\documentstyle[12pt,epsf]{article}
\begin{document}

\begin{titlepage}
\baselineskip 20pt

\title{
\begin{flushright}
{\large\bf MZ-TH/96-32\\
WU-B 96-44\\
December 1996}\\[5mm]
\end{flushright}
Heavy Baryon Transitions in\\
a Relativistic Three-Quark Model}

\author{M.A. Ivanov, V.E. Lyubovitskij\\
Bogoliubov Laboratory of Theoretical Physics,\\
Joint Institute for Nuclear Research,\\
141980 Dubna (Moscow region), Russia\\
\\
J.G. K\"{o}rner\\
Johannes Gutenberg-Universit\"{a}t, \\
Institut f\"{u}r Physik, \\
D-55099 Mainz, Germany\\
\\
P. Kroll\\
Fachbereich Physik, Universit\"{a}t Wuppertal,\\
D-42097 Wuppertal 1, Germany\\
\\
}

\maketitle

\abstract
\baselineskip 20pt
Exclusive semileptonic decays of bottom and charm baryons are considered
within a relativistic three-quark model with a Gaussian shape
for the baryon-three-quark vertex and standard quark propagators.
We calculate the baryonic Isgur-Wise functions, decay rates and
asymmetry parameters.

\begin{center}
\end{center}

\end{titlepage}

\section{Introduction}
\baselineskip 20pt

The investigation of semileptonic (s.l.) decays of heavy hadrons allows
one to determine
the unknown Cabibbo-Kabayashi-Maskawa (CKM) matrix elements, i.e.
$V_{bc}$ and $V_{bu}$ in bottom meson and baryon decays.
These play a fundamental role in the physics of weak
interactions. The CKM matrix elements can be extracted from the inclusive
s.l. width of heavy hadron~\cite{Shifman} or decay spectra~\cite{Blok} and
from the exclusive differential rates of $B\to D^\star l\nu$,
$\Lambda_b\to\Lambda_c l\nu$, ..., extrapolated to the point of zero
recoil~\cite{Shifman}, \cite{Neubert}-\cite{Koerner1}.
Other characteristics of semileptonic decays (momentum dependence of
transition form factors, exclusive decay rates, asymmetry parameters and etc.)
are also important for our understanding of the heavy hadron structure.

From a modern point of view the appropriate theoretical framework for the
analysis of hadrons containing a single heavy quark is the Heavy Quark
Effective Theory (HQET)~\cite{Politzer}-\cite{Hussain} based on a systematic
$1/m_Q$-expansion of the QCD Lagrangian. The leading order of
the HQET-expansion, when the heavy quark mass goes to infinity, corresponds
to the case of Heavy Quark Symmetry (or Isgur-Wise symmetry) \cite{Isgur1}.
Due to the Isgur-Wise (IW) symmetry the structure of weak currents
of low-lying baryons is simplified. The form factors of these transitions
are expressed through a few
universal functions. Unfortunately, HQET can give predictions only for
the normalization of the form factors at zero recoil.
Once one moves away from the zero recoil point
one has to take recource to full nonperturbative calculations.

This paper focuses on exclusive s.l. decays of the ground state
bottom and charm baryons. Recently, the activity in this field has
started to make contact with experiment due to the observation
of the CLEO Collaboration \cite{CLEO} of the heavy-to-light s.l. decay
mode $\Lambda_c^+\to\Lambda e^+\nu_e$.
Also the ALEPH \cite{ALEPH} and OPAL \cite{OPAL} Collaborations expect
to observe the exclusive mode $\Lambda_b\to\Lambda_c\ell\nu$
in the near future. Therefore, a theoretical study of the s.l. decays of
heavy baryons seems to be very important.

In \cite{Aniv1,Aniv2} a model for QCD bound states composed of
light and heavy quarks was proposed. The model is
the Lagrangian formulation of the NJL model with separable interaction
\cite{Goldman,Gross} but its advantage consists in the possibility of
studying baryons as relativistic systems of three quarks.
The general framework was developed  for light mesons \cite{Aniv1,Aniv2} and
baryons \cite{Aniv2,PSI}, and also for heavy-light hadrons \cite{Manchester}.
Particularly, in ref.~\cite{Aniv1,Aniv2} the pion weak decay constant,
the two-photon decay width, as well as the form factor of the
$\gamma^*\pi^0\to\gamma$ transition, the pion charge form factor, and
the strong $\pi NN$ form factor have been calculated and good
agreement with the data has been achieved with three parameters. Two of
the parameters are range parameters characterizing the size of mesons
and baryons. The remaining parameter is the constituent quark mass.
In ref.~\cite{PSI} the approach developed in \cite{Aniv1,Aniv2} was
applied to a calculation of the electromagnetic form factors
of nucleons. Some preliminary results on s.l. decays of heavy-light
baryons were already presented in \cite{Manchester}.

The purpose of the present work is to give a description of the properties of
baryons containing a single heavy quark within the framework proposed in
\cite{Aniv1,Aniv2} and developed in \cite{PSI,Manchester}.
Namely, we report the calculation of observables in semileptonic decays of
bottom and charm baryons: Isgur-Wise functions, asymmetry parameters,
decay rates and distributions.

\section{Model}

We start with a brief review of our approach \cite{Aniv1,Aniv2}
based on interaction Lagragians coupling hadrons with constituent quarks
and {\it vice versa}. It was found
\cite{Aniv1,Aniv2,PSI,Manchester} that this approach succesfully
describes low-energy hadronic properties like decay constants,
form factors, etc.  Here we are going to apply this approach
to the calculation of baryonic observables when the baryons contain a
heavy (b or c) quark.

Let  $y_i$ (i=1,2,3) be the spatial 4-coordinates of quarks with
masses $m_i$, respectively.  They are expressed through the center of mass
coordinate $(x)$ and relative Jacobi coordinates $(\xi_1, ...)$ as
\begin{eqnarray}\label{cmf}
& &y_1=x-3\xi_1\,\,\frac{m_2+m_3}{\sum\limits_{i}m_i}
\nonumber\\
& &y_2=x+3\xi_1\,\,\frac{m_1}{\sum\limits_{i}m_i}-
2\xi_2\sqrt{3}\,\,\frac{m_3}{m_2+m_3}
\\
& &y_3=x+3\xi_1\,\,\frac{m_1}{\sum\limits_{i}m_i}+
2\xi_2\sqrt{3}\,\,\frac{m_2}{m_2+m_3}
\nonumber\\
& &\nonumber\\
\mbox{where}& &x=\frac{\sum\limits_{i}m_iy_i}{\sum\limits_{i}m_i},
\hspace{0.5cm}
\xi_1=\frac{1}{3}\,\,\bigg(\frac{m_2y_2+m_3y_3}{m_2+m_3}-y_1\biggr),
\hspace{0.5cm}
\xi_2=\frac{y_3-y_2}{2\sqrt{3}}.
\nonumber
\end{eqnarray}
We assume that the momentum distribution of the constituents inside
a baryon is modeled by an effective relativistic vertex function
$F\left(\frac{1}{18}\sum\limits_{i<j}(y_i-y_j)^2\right)$
which depends on the sum of relative configuration space coordinates only.
Its fall-of is sufficient to guarantee ultraviolet
convergence of
matrix elements. At the same time the vertex function is a phenomenological
description of the long distance QCD interactions between quarks and gluons.

Then the general form of the interaction Lagrangian of baryons with quarks
is written as
\begin{eqnarray}\label{strong}
{\cal L}_B^{\rm int}(x)&=&g_B\bar B(x)
\hspace*{-.1cm}\int \hspace*{-.1cm}dy_1\hspace*{-.1cm}\int
\hspace*{-.1cm}dy_2\hspace*{-.1cm}\int \hspace*{-.1cm}dy_3 \,
\delta\left(x-\frac{\sum\limits_i m_iy_i}{\sum\limits_i m_i}\right)
F\left(\frac{1}{18}\sum\limits_{i<j}(y_i-y_j)^2\right)\nonumber \\
& &\nonumber\\
&\times&J_B(y_1,y_2,y_3)+h.c.
\end{eqnarray}
with $J_B(y_1,y_2,y_3)$ being the 3-quark current with quantum numbers
of a baryon $B$:
\begin{equation}\label{current}
J_B(y_1,y_2,y_3)=\Gamma_1 q^{a_1}(y_1)q^{a_2}(y_2)C\Gamma_2 q^{a_3}(y_3)
\varepsilon^{a_1a_2a_3}.
\end{equation}
Here $\Gamma_{1(2)}$ are strings of Dirac matrices,
$C=\gamma^0\gamma^2$ is
the charge conjugation matrix, and $a_i$ are the color indices.

The choice of baryonic currents depends on two different
cases:\\
a) light baryons composed from $u, d, s$ quarks,\\
b) heavy-light baryons with a single heavy quark $b$ or $c$.

In the case of light baryons we shall work in the limit of isospin
invariance by assuming that the masses of $u$ and $d$ quarks are equal
to each other, i.e.
$m_u=m_d=m$. The breaking of SU(3) symmetry is taken into account
via a difference of strange and nonstrange quark masses $m_s-m\neq 0$.
Thus, for baryons composed either of $u$ or $d$ quarks
(nucleons, $\Delta$-isobar) or of $s$ quarks ($\Omega$-hyperon)
the coordinates of quarks may be written as
\begin{eqnarray}
& &y_1=x-2\xi_1 \hspace{0.5cm} y_2=x+\xi_1-\xi_2\sqrt{3}
\hspace{0.5cm} y_3=x+\xi_1+\xi_2\sqrt{3} \nonumber
\end{eqnarray}
If a light baryon contains a single strange quark with mass $m_s$ and two
nonstrange quarks ($u$ or $d$) with a mass $m$ each
 as in $\Lambda$ and $\Sigma$-hyperons one gets
\begin{eqnarray}
& &y_1=x-6\xi_1\,\,\frac{m}{2m+m_s}
\nonumber\\
& &y_2=x+3\xi_1\,\,\frac{m_s}{2m+m_s}-\xi_2\sqrt{3},
\hspace*{1cm}
y_3=x+3\xi_1\,\,\frac{m_s}{2m+m_s}+\xi_2\sqrt{3}\nonumber
\end{eqnarray}
\noindent where $y_1$ is the coordinate of the strange quark and
$y_2$ and $y_3$ are the coordinates of nonstrange quarks.

For a baryon with two strange quarks and a single
nonstrange quark
(as e.g. in the $\Xi-$hyperons) one obtains
\begin{eqnarray}
& &y_1=x-6\xi_1\,\,\frac{m_s}{2m_s+m} \nonumber\\
& &y_2=x+3\xi_1\,\,\frac{m}{2m_s+m}-\xi_2\sqrt{3},
\hspace*{1cm}
y_3=x+3\xi_1\,\,\frac{m}{2m_s+m}+\xi_2\sqrt{3}
\nonumber
\end{eqnarray}
\noindent where $y_1$ now is the coordinate of the nonstrange quark and
$y_2$ and $y_3$ are the coordinates of the strange quarks.

The spin-flavor structure of light baryonic currents with quantum
numbers $J^P=\frac{1}{2}^+$ and $J^P=\frac{3}{2}^+$ has been studied
in detail in the papers \cite{Ioffe}-\cite{Lubovit}.
It was shown that there are two possibilities to choose the baryonic
currents with $J^P=\frac{1}{2}^+$:
\begin{eqnarray}\label{vect-l}
\mbox{\it \underline{vector variant}} \hspace*{.3cm}
J_B^V(y_1,y_2,y_3)=\gamma^\mu\gamma^5 q^{a_1}(y_1)q^{a_2}(y_2)
C\gamma_\mu q^{a_3}(y_3)\varepsilon^{a_1a_2a_3}
\end{eqnarray}
\begin{eqnarray}\label{tens-l}
\mbox{\it \underline{tensor variant}}\hspace*{.3cm}
J_B^T(y_1,y_2,y_3)=\sigma^{\mu\nu}\gamma^5 q^{a_1}(y_1)q^{a_2}(y_2)
C\sigma_{\mu\nu} q^{a_3}(y_3)\varepsilon^{a_1a_2a_3}
\end{eqnarray}
Both of these forms have been used in \cite{Lubovit,Diquark} for studying
the electromagnetic and strong properties of light baryons. It was shown that
the {\it tensor variant} is more suitable for the
description of the data. For this reason we will use the tensor current
in the approach developed in this paper. For convenience the tensor
current can be transformated into a sum of {\it pseudoscalar}
($\Gamma_1=I, \Gamma_2=\gamma_5$ in Eq.~(\ref{current})) and
{\it scalar currents} ($\Gamma_1=\gamma_5, \Gamma_2=I$ in
Eq.~(\ref{current})) using the Fierz transformations:
\begin{eqnarray}
(\sigma^{\mu\nu}\gamma^5)_{i_1i_2}(C\sigma_{\mu\nu})_{i_3i_4}=
-2[I_{i_1i_2}(C\gamma_5)_{i_3i_4}+ \gamma^5_{i_1i_2}C_{i_3i_4}] +
 4[I_{i_1i_4}(C\gamma_5)_{i_3i_2}+ \gamma^5_{i_1i_4}C_{i_3i_2}]
\nonumber
\end{eqnarray}
\noindent For example, a {\it tensor current} for the proton
\begin{eqnarray}
J_p^T(y_1,y_2,y_3)=\sigma^{\mu\nu}\gamma^5 d^{a_1}(y_1)u^{a_2}(y_2)
C\sigma_{\mu\nu} u^{a_3}(y_3)\varepsilon^{a_1a_2a_3}\nonumber
\end{eqnarray}
\noindent written in $S+P$ form becomes
\begin{eqnarray}\label{curr-pr}
J_p^T(y_1,y_2,y_3)=4[u^{a_1}(y_3)u^{a_2}(y_2)
C\gamma_5 d^{a_3}(y_1)+ \gamma_5 u^{a_1}(y_3)u^{a_2}(y_2)
C d^{a_3}(y_1)]\varepsilon^{a_1a_2a_3}\nonumber
\end{eqnarray}
\noindent in the Fierz transformed form.
After exchanging the variables $y_1\leftrightarrow y_3$ in the interaction
Lagrangian of the proton with quarks we have
\begin{eqnarray}\label{lagr-pr}
{\cal L}_P^{\rm int, T}(x)&=&4g_p^T\bar p(x)
\hspace*{-.1cm}\int \hspace*{-.1cm}dy_1\hspace*{-.1cm}\int
\hspace*{-.1cm}dy_2\hspace*{-.1cm}\int \hspace*{-.1cm}dy_3 \,
\delta\left(x-\frac{\sum\limits_i m_iy_i}{\sum\limits_i m_i}\right)
F\left(\frac{1}{18}\sum\limits_{i<j}(y_i-y_j)^2\right)\nonumber \\
& &\nonumber\\
&\times&[u^{a_1}(y_1)u^{a_2}(y_2)
C\gamma_5 d^{a_3}(y_3)+ \gamma_5 u^{a_1}(y_1)u^{a_2}(y_2)
C d^{a_3}(y_3)]\varepsilon^{a_1a_2a_3}+h.c.\nonumber
\end{eqnarray}

\begin{figure}[htbp]
\begin{center}
{\bf Table 1. Three-Quark Currents of Light Baryons\\}
\end{center}
\def\arraystretch{3.0}
\begin{center}
\begin{tabular}{|c|c|}
\hline
\hline
Baryon & Three-Quark Current\\
\hline
Proton & $J_p^T(y_1,y_2,y_3)=
[u^a(y_1) u^b(y_2)C\gamma^5 d^c(y_3)
+\gamma^5u^a(y_1) u^b(y_2)Cd^c(y_3)] \varepsilon^{abc}$\\
\hline
Neutron & $J_n^T(y_1,y_2,y_3)=
[d^a(y_1) d^b(y_2)C\gamma^5 u^c(y_3)
+\gamma^5d^a(y_1) d^b(y_2)Cu^c(y_3)] \varepsilon^{abc}$\\
\hline
$\Xi^-$-hyperon &
$J_{\Xi^-}^T(y_1,y_2,y_3)=
[s^a(y_1) s^b(y_2)C\gamma^5 d^c(y_3)
+\gamma^5s^a(y_1) s^b(y_2)Cd^c(y_3)] \varepsilon^{abc}$\\
\hline
$\Lambda^0$-hyperon &
$J_{\Lambda^0}^T(y_1,y_2,y_3)=
[s^a(y_1) u^b(y_2)C\gamma^5 d^c(y_3)
+\gamma^5s^a(y_1) u^b(y_2)Cd^c(y_3)] \varepsilon^{abc}$\\
\hline
\hline
\end{tabular}
\end{center}
\end{figure}
Table 1 contains a set of {\it tensor currents} for nucleons,
$\Lambda^0$ and $\Xi^-$-hyperons in the $S+P$ form
which will be used in our calculations.

Next we turn to the discussion of heavy-light baryonic currents.
Suppose that the heavy-quark mass is much larger than the light-quark masses
$(m_Q\gg m_{q_1},m_{q_2})$. From Eq.~(\ref{cmf}) one then obtains:
\begin{eqnarray}
& &y_1=y_Q=x\nonumber\\
& &y_2=y_{q_1}=x+3\xi_1-2\xi_2\sqrt{3}\,\,\frac{m_{q_2}}{m_{q_1}+m_{q_2}}
\,\,\,\,\,\mbox{and}\,\,\,\,\,
y_3=y_{q_2}=x+3\xi_1+2\xi_2\sqrt{3}\,\,\frac{m_{q_1}}{m_{q_1}+m_{q_2}}
\nonumber
\end{eqnarray}
\noindent where $y_1$ is the coordinate of heavy quark, and
$y_2$ and $y_3$ are the coordinates of light quarks $q_1$ and $q_2$.

It is convenient to transform the Jacobi coordinates of Eq.(\ref{cmf})
to remove the light-quark mass dependence
\begin{eqnarray}
\xi_1&\to& \xi_1-\frac{\xi_2}{\sqrt{3}}\,\,
\frac{m_{q_1}-m_{q_2}}{m_{q_1}+m_{q_2}}
\nonumber\\
\xi_2&\to& \xi_2
\nonumber
\end{eqnarray}
Then we have
\begin{eqnarray}
y_1=y_Q=x, \hspace*{1cm} y_2=y_{q_1}=x+3\xi_1-\xi_2\sqrt{3},
           \hspace*{1cm} y_3=y_{q_2}=x+3\xi_1+\xi_2\sqrt{3}
\nonumber
\end{eqnarray}

The problem of the spin-flavor structure of heavy-light baryonic currents was
analyzed in ref.~\cite{Shuryak}-\cite{Groote}.
It was shown that, in the static limit $\vec p_Q\to 0$
(this is equivalent to the heavy quark limit $m_Q\to\infty$),
$\Lambda$-type baryons ($\Lambda_Q$, $\Xi_Q$) containing
a light diquark system with zero spin may be described by either of the
following nonderivative three-quark currents
\begin{eqnarray}
& &J_{\Lambda_{h_Q}}^P=\varepsilon^{abc}h_Q^au^bC\gamma^5d^c,\,\,\,\,\,
J_{\Lambda_{h_Q}}^A=\varepsilon^{abc}h_Q^au^b
C\gamma^0\gamma^5d^c\nonumber
\end{eqnarray}
\noindent where $h_Q$ denotes the effective static field of the heavy quark.

In the same vein there are two currents for $\Omega$-type baryons
($\Omega_Q$, $\Sigma_Q$ and $\Omega^\star_Q$, $\Sigma^\star_Q$)
containing a light diquark system with spin 1
\begin{eqnarray}
& &J_{\Omega_{h_Q}}^{V}=\varepsilon^{abc}\vec\gamma\gamma_5h_Q^a
s^bC\vec\gamma s^c,\,\,\,\,\,
J_{\Omega^\star_{h_Q}}^{V;k}=\varepsilon^{abc}[h_Q^a
s^bC\gamma^k s^c+\frac{1}{3}\gamma^k\vec\gamma
h_Q^as^bC\vec\gamma s^c]\nonumber\\
& &\nonumber\\
& &J_{\Omega_{h_Q}}^{T}=\varepsilon^{abc}\vec\gamma\gamma_5h_Q^a
s^bC\gamma^0\vec\gamma s^c,\,\,\,\,\,
J_{\Omega^\star_{h_Q}}^{T;k}=\varepsilon^{abc}
[h_Q^as^bC\gamma^0\gamma^ks^c+\frac{1}{3}\gamma^k\vec\gamma
h_Q^as^bC\gamma^0\vec\gamma s^c]\nonumber
\end{eqnarray}
\noindent where $k=1,2,3$, $(\gamma^k)^2=-3$.
The currents $J_{\Omega^\star_{h_Q}}^{I;k} \,(I=V,T)$ satisfy the
spin-$3/2$ Rarita-Schwinger condition
$\gamma^k J_{\Omega^\star_{h_Q}}^{I;k}=0$.
In this paper we work with Lorentz-covariant representations
of the HQET heavy-light currents
mentioned above \cite{Shuryak}-\cite{Groote}.

Our currents are listed below
\begin{eqnarray}
& &\mbox{\underline{\it pseudoscalar variant}} \hspace*{.5cm}
J_{\Lambda_{h_Q}}^P\to
J_{\Lambda_Q}^P=\varepsilon^{abc}Q^au^bC\gamma^5d^c\label{pseud-h}\\
& &\nonumber\\
& &\mbox{\underline{\it axial variant}} \hspace*{1.8cm}
J_{\Lambda_{h_Q}}^A \to
J_{\Lambda_Q}^A=\varepsilon^{abc}\gamma_\mu Q^au^b
C\gamma^\mu\gamma^5d^c\label{axial-h}\\
& &\nonumber\\
& &\mbox{\underline{\it vector variant}}\hspace*{1.6cm}
J_{\Omega_{h_Q}}^{V} \to
J_{\Omega_Q}^{V}=\varepsilon^{abc}\gamma_\mu\gamma^5Q^a
s^bC\gamma^\mu s^c\label{vect1-h}\\[3mm]
& &\hspace*{4.2cm}J_{\Omega^\star_{h_Q}}^{V;k} \to
J_{\Omega^\star_Q}^{V;\mu} + J_{\Omega^\star_Q}^{(\perp)V;\mu}
\nonumber\\[3mm]
& &\hspace*{4.2cm}J_{\Omega^\star_Q}^{V;\mu}=\varepsilon^{abc}Q^a
s^bC\gamma^\mu s^c\label{vect2-h}\\[3mm]
& &\hspace*{4.2cm}J_{\Omega^\star_Q}^{(\perp)V;\mu}=-\frac{1}{4}
\varepsilon^{abc}\gamma^\mu\gamma_\nu Q^as^bC\gamma^\nu s^c\nonumber\\[3mm]
& &\mbox{\underline{\it tensor variant}}\hspace*{1.6cm}
J_{\Omega_{h_Q}}^{T} \to
J_{\Omega_Q}^{T}=\varepsilon^{abc}\sigma_{\mu\nu}\gamma_5Q^a
s^bC\sigma^{\mu\nu}s^c,\label{tens1-h}\\[3mm]
& &\hspace*{4.2cm}J_{\Omega^\star_{h_Q}}^{T;k} \to
J_{\Omega^\star_Q}^{T;\mu} + J_{\Omega^\star_Q}^{(\perp)T;\mu}
\nonumber\\[3mm]
& &\hspace*{4.2cm}J_{\Omega^\star_Q}^{T;\mu}=
-i\varepsilon^{abc}\gamma_\nu Q^a
s^bC\sigma^{\mu\nu} s^c\label{tens2-h}\\[3mm]
& &\hspace*{4.2cm}
J_{\Omega^\star_Q}^{(\perp)T;\mu}=\frac{i}{4}\varepsilon^{abc}
\gamma^\mu\gamma_\alpha\gamma_\nu Q^as^bC\sigma^{\alpha\nu} s^c\nonumber
\end{eqnarray}
\noindent The currents $J_{\Omega^\star_Q}^{(\perp)I;\mu} (I=V,T)$
are orthogonal to the corresponding baryon field with spin $3/2$:
$\bar\Omega^{\star \mu}_Q \cdot J_{\Omega^\star_Q}^{(\perp)I;\mu} = 0$ and
can, therefore, be omitted in the interaction Lagrangian (\ref{strong}).
Thus, for heavy-light baryons with spin $3/2$ we use
the currents $J_{\Omega^\star_Q}^{V}$ (\ref{vect2-h}) and
$J_{\Omega^\star_Q}^{T}$ (\ref{tens2-h}).

In Table 2 we give the quark content, the quantum numbers
(spin-parity $J^P$, spin $S_{qq}$ and isospin
$I_{qq}$ of light diquark) and the experimental (when available) and
theoretical mass spectrum of heavy baryons
\cite{DESY,PDG} which will be analyzed in this paper.
Square brackets $[...]$ and round brackets $\{...\}$ denote antisymmetric
and symmetric flavor and spin combinations
of the light degrees of freedom.

\begin{figure}[htbp]
\begin{center}
{\bf Table 2. Quantum Numbers of Heavy-Light Baryons\\}
\end{center}
\vspace*{0.2cm}
\def\arraystretch{2.0}
\begin{center}
\begin{tabular}{|c|c|c|c|c|}
\hline
\hline
Baryon & Quark Content & $J^P$ & $(S_{qq}, I_{qq})$ & Mass (GeV)\\
\hline
\hline
$\Lambda_c^+$ &  c[ud] & ${\frac{1}{2}}^+$ & (0,0) & 2.285\\
\hline
$\Xi_c^+$ &  c[us] & ${\frac{1}{2}}^+$ & (0,1/2) & 2.466\\
\hline
$\Sigma_c^{++}$ &  c\{uu\} & ${\frac{1}{2}}^+$ & (1,1) & 2.453\\
\hline
$\Omega_c^0$ &  c\{ss\} & ${\frac{1}{2}}^+$ & (1,0) & 2.719\\
\hline
$\Sigma_c^{\star ++}$ &  c\{uu\} & ${\frac{3}{2}}^+$ & (1,1) & 2.510\\
\hline
$\Omega_c^{\star 0}$ &  c\{ss\} & ${\frac{3}{2}}^+$ & (1,0) & 2.740\\
\hline
$\Lambda_b^0$ &  b[ud] & ${\frac{1}{2}}^+$ & (0,0) & 5.640\\
\hline
$\Xi_b^+$ &  b[us] & ${\frac{1}{2}}^+$ & (0,1/2) & 5.800\\
\hline
$\Sigma_b^+$ &  b\{uu\} & ${\frac{1}{2}}^+$ & (1,1) & 5.820\\
\hline
$\Omega_b^-$ &  b\{ss\} & ${\frac{1}{2}}^+$ & (1,0) & 6.040\\
\hline
\end{tabular}
\end{center}

\unitlength=0.50mm
\special{em:linewidth 0.4pt}
\linethickness{0.4pt}
\begin{center}
\hspace*{-1.6cm}
\begin{picture}(127.00,99.00)
\put(70.00,52.00){\oval(48.00,16.00)[]}
\put(45.00,52.00){\circle*{5.20}}
\put(95.00,52.00){\circle*{5.20}}
\put(95.00,52.00){\line(1,0){25.00}}
\put(120.00,53.00){\line(-1,0){25.00}}
\put(95.00,51.00){\line(1,0){25.00}}
\put(45.00,51.00){\line(-1,0){25.00}}
\put(20.00,52.00){\line(1,0){25.00}}
\put(45.00,53.00){\line(-1,0){25.00}}
\put(13.00,52.00){\makebox(0,0)[cc]{{\bf B$_q$}}}
\put(127.00,52.00){\makebox(0,0)[cc]{{\bf B$_q$}}}
\put(65.00,73.00){\makebox(0,0)[cc]{{\bf q}}}
\put(65.00,55.00){\makebox(0,0)[cc]{{\bf q}}}
\put(65.00,40.00){\makebox(0,0)[cc]{{\bf q}}}
\put(32.00,52.05){\line(2,1){9.00}}
\put(32.00,52.05){\line(2,-1){9.00}}
\put(108.00,52.05){\line(2,1){9.00}}
\put(108.00,52.05){\line(2,-1){9.00}}
\put(70.00,50.00){\oval(50.50,35.50)[t]}
\put(68.00,67.80){\line(4,1){9.00}}
\put(68.00,67.80){\line(4,-1){9.00}}
\put(68.00,60.10){\line(4,1){9.00}}
\put(68.00,60.10){\line(4,-1){9.00}}
\put(68.00,44.10){\line(4,1){9.00}}
\put(68.00,44.10){\line(4,-1){9.00}}
\end{picture}
\hspace*{1.1cm}
\begin{picture}(127.00,99.00)
\put(70.00,52.00){\oval(48.00,16.00)[]}
\put(45.00,52.00){\circle*{5.20}}
\put(95.00,52.00){\circle*{5.20}}
\put(95.00,52.00){\line(1,0){25.00}}
\put(120.00,53.00){\line(-1,0){25.00}}
\put(95.00,51.00){\line(1,0){25.00}}
\put(45.00,51.00){\line(-1,0){25.00}}
\put(20.00,52.00){\line(1,0){25.00}}
\put(45.00,53.00){\line(-1,0){25.00}}
\put(13.00,52.00){\makebox(0,0)[cc]{{\bf B$_Q$}}}
\put(127.00,52.00){\makebox(0,0)[cc]{{\bf B$_Q$}}}
\put(65.00,73.00){\makebox(0,0)[cc]{{\bf Q}}}
\put(65.00,55.00){\makebox(0,0)[cc]{{\bf q}}}
\put(65.00,40.00){\makebox(0,0)[cc]{{\bf q}}}
\put(32.00,52.05){\line(2,1){9.00}}
\put(32.00,52.05){\line(2,-1){9.00}}
\put(108.00,52.05){\line(2,1){9.00}}
\put(108.00,52.05){\line(2,-1){9.00}}
\put(70.00,50.00){\oval(50.50,36.50)[t]}
\put(70.00,50.00){\oval(50.50,36.)[t]}
\put(70.00,50.00){\oval(50.50,35.50)[t]}
\put(70.00,50.00){\oval(50.50,35.)[t]}
\put(70.00,50.00){\oval(50.50,34.5)[t]}
\put(68.00,67.80){\line(4,1){9.00}}
\put(68.00,67.80){\line(4,-1){9.00}}
\put(68.00,60.10){\line(4,1){9.00}}
\put(68.00,60.10){\line(4,-1){9.00}}
\put(68.00,44.10){\line(4,1){9.00}}
\put(68.00,44.10){\line(4,-1){9.00}}
\end{picture}
\vspace*{-0.5cm}
\mbox{{\bf Fig.1a}. Light baryon mass operator.}
\ \ \
\mbox{{\bf Fig.1b}. Heavy-light baryon mass operator.}
\end{center}
\end{figure}
The Lagrangian that  describes the interaction of
$\Lambda^0_b$ with $b, u, d$ - quarks
is then written as
\begin{eqnarray}
{\cal L}_{\Lambda^0_b}^{\rm int}(x)&=&g_B\bar \Lambda^0_b(x)
\Gamma_1 b^a(x)\int d\xi_1 \int d\xi_2 F\left(\xi_1^2+\xi_2^2\right)\\
&\times&u^b(x+3\xi_1-\xi_2\sqrt{3})C\Gamma_2d^c(x+3\xi_1+\xi_2\sqrt{3})
\varepsilon^{abc}+h.c.
\nonumber
\end{eqnarray}
\noindent where
\[\Gamma_1\otimes C\Gamma_2=\left\{
\begin{array}{ll}
I\otimes C\gamma^5
&\,\,\,\,\mbox{pseudoscalar current}\\[3mm]
\gamma_\mu \otimes C\gamma^\mu\gamma^5
&\,\,\,\,\mbox{axial current}
\end{array}\right. \]

The vertex form factor $F(\xi^2_1+\xi^2_2)$ characterizes the
distribution of $u$ and $d$ quarks inside the $\Lambda^0_b$ baryon.
The Fourier-transform of the vertex form factor is defined as
\begin{eqnarray}\label{ff}
F(\xi^2_1+\xi^2_2)=\int\frac{d^4k_1}{(2\pi)^4}\int\frac{d^4k_2}{(2\pi)^4}
\exp(-ik_1\xi_1-ik_2\xi_2)F(k^2_1+k^2_2)
\end{eqnarray}

Next we discuss the model parameters. First, there are the baryon-quark
coupling constants and the vertex function in the Lagrangian (\ref{strong}).
The coupling constants are calculated from {\it the compositeness condition}
(see, ref.~\cite{QCM}), i.e. the renormalization constant of the
baryon wave function is set equal to zero,
$Z_B=1-g_B^2\Sigma^\prime_B(M_B)=0$, with $\Sigma_B$ being the baryon mass
operator (see, Fig.1a for light baryons and Fig.1b for heavy baryons).
Actually, the compositeness condition is equivalent to the normalization
of the elastic form factors to one at zero momentum transfer.
This may be readily seen from the Ward indentity which relates the vertex
function with the mass operator on mass shell. We have

\begin{eqnarray}\label{LSI}
\left.\Lambda^\mu_{B\to B\gamma}(p,p)\right|_{\,\not p=M_B}=
\left. g_B^2 \frac{\partial\Sigma_B(p)}{\partial p^\mu}
\right|_{\,\not p=M_B}=
\left.\gamma^\mu g_B^2\Sigma^\prime_B(p)\right|_{\,\not p=M_B}
\end{eqnarray}
where the vertex function is related to the baryon elastic form factor by

\begin{eqnarray}\label{FB0}
\Lambda^\mu_{B\to B\gamma}(p,p)=\gamma^\mu F_B(0).
\end{eqnarray}
From this the normalization of the
form factor mentioned above immediately follows.

The vertex function is an arbitrary function except that it should make the
Feynman diagrams ultraviolet finite, as we have mentioned above.
In the papers \cite{Aniv1,Aniv2} we have found that the basic physical
observables of pion and nucleon low-energy physics depend only weakly on
the choice of the vertex functions.
In this paper we choose a Gaussian vertex function for simplicity.
In Minkowski space we write
\begin{eqnarray}
F(k^2_1+k^2_2)&=&\exp\biggl(\frac{k^2_1+k^2_2}{\Lambda_B^2}\biggr)
\nonumber
\end{eqnarray}
where $\Lambda_B$ is the Gaussian range parameter which may be
related to the size of a baryon. Note that all calculations
are done in the Euclidean region ($k^2_i=-k^2_{i E}$) where the
above vertex function decreases very rapidly.
It was found in \cite{PSI} that for nucleons
$(B=N)$ the value $\Lambda_N=1.25$ GeV gives a good description of the
nucleon's static characteristics (magnetic moments, charge radii)
and its form factors in the space-like region for $Q^2$ up
to 1 GeV$^2$. In this work we will use the value
$\Lambda_{B_q}\equiv\Lambda_N=1.25$ GeV
for light baryons and take the value $\Lambda_{B_Q}$ for the heavy-light
baryons as an adjustable parameter.

For light quark propagator with a mass $m_q$ we shall use
the standard form of the free fermion propagator
\begin{eqnarray}\label{Slight}
<0|{\rm T}(q(x)\bar q(y))|0>=
\int{d^4k\over (2\pi)^4i}e^{-ik(x-y)}S_q(k), \,\,\,\,\,\,
S_q(k)={1\over m_q-\not\! k}
\end{eqnarray}
For the heavy quark propagator we will use the leading term in the
inverse mass expansion. Suppose $p=M_{B_Q}v$ is the heavy baryon momentum.
We introduce the parameter $\bar\Lambda_{\{q_1q_2\}}=M_{\{Qq_1q_2\}}-m_Q$
which is the difference between the heavy baryon mass
$M_{\{Qq_1q_2\}}\equiv M_{B_Q}$
and the heavy quark mass. Keeping in mind that the vertex function
falls off sufficiently fast such that the condition
$|k|<<m_Q$ holds where $k$ is the virtual momentum of light quarks, one has
\begin{eqnarray}\label{Sheavy}
& &S_Q(p+k)={1\over m_Q - (\not\! p \,\, + \not\! k)}=
\frac {m_Q+M_{B_Q}\not\! v+\not\! k} {m_Q^2-M_{B_Q}^2-2M_{B_Q}vk-k^2}
=S_v(k,\bar\Lambda_{\{q_1q_2\}})+O\biggl(\frac{1}{m_Q}\biggr)\nonumber\\[5mm]
& &S_v(k,\bar\Lambda_{\{q_1q_2\}}) = - \frac{(1+\not\! v)}
{2(v\cdot k + \bar\Lambda_{\{q_1q_2\}})}
\end{eqnarray}
In what follows we will assume that
$\bar\Lambda\equiv\bar\Lambda_{uu}=\bar\Lambda_{dd}=\bar\Lambda_{du}$,
$\bar\Lambda_{s}\equiv\bar\Lambda_{us}=\bar\Lambda_{ds}$. Thus there are three
independent parameters: $\bar\Lambda$, $\bar\Lambda_{s}$, and
$\bar\Lambda_{ss}$.

A drawback of our approach is the lack of confinement. This
can in principle be corrected by changing the analytic properties
of the light-quark propagator.
We leave the investigation of this possibility for future studies.
For the time being we shall
avoid the appearance of unphysical imaginary parts in the Feynman diagrams
by postulating the following condition:
the baryon mass must be less than the sum of constituent quark masses
$M_B<\sum\limits_i m_{q_i}$.

In the case of heavy-light baryons the restriction
$M_B<\sum\limits_i m_{q_i}$ implies that the parameter
$\bar\Lambda_{\{q_1q_2\}}$ must be less than the
sum of light quark masses $\bar\Lambda_{\{q_1q_2\}} < m_{q_1}+m_{q_2}$. The
last constraint serves as the upper limit for our choices of the parameter
$\bar\Lambda_{\{q_1q_2\}}$.

Thus, there are three sets of adjustable parameters in our model:
the constituent light quark masses $m_q$ ($m=m_u=m_d$ and $m_s$), the range
cutoff parameters $\Lambda_B$ ($\Lambda_{B_q}$ and $\Lambda_{B_Q}$)
and a set of $\bar\Lambda_{\{q_1q_2\}}$ subsidiary parameters: $\bar\Lambda$,
$\bar\Lambda_s$ and $\bar\Lambda_{\{ss\}}$. The parameters $m$=420 MeV and
$\Lambda_{B_q}$=1.25 GeV were fixed in ref.~\cite{PSI} from a best
fit to the data on electromagnetic properties of nucleons.
The parameters $\Lambda_{B_Q}$, $m_s$, $\bar\Lambda$ are determined in
this paper
from the analysis of the $\Lambda^+_c\to\Lambda^0+e^+ +\nu_e$ decay data.
The following values are obtained: $\Lambda_Q$=2.5 GeV, $m_s$=570 MeV and
$\bar\Lambda$=710 MeV.

The parameters $\bar\Lambda_s$ and $\bar\Lambda_{\{ss\}}$ cannot be adjusted
at present since at present
there are no experimental data on the decays of heavy-light
baryons containing one or two strange quarks.

\vspace*{1.5cm}
\section{Matrix Elements of Semileptonic Decays of Bottom
and Charm Baryons}

In our model the semileptonic decays of bottom and charm baryons are described
by the standard triangle quark diagram (Fig.2).
The matrix elements describing heavy-to-heavy ~($b\to c$) and
heavy-to-light ($c\to s$) transitions can be written as

$\bullet$ $b\to c$ transition
\begin{eqnarray}\label{BC-trans}
\bar u(v^\prime)M_\Gamma(v,v^\prime)u(v)&=&
g_{B_b}g_{B_c}\hspace*{-0.1cm}\int\hspace*{-0.1cm}\frac{d^4k}{\pi^2i}
\hspace*{-0.1cm}\int\hspace*{-0.1cm}\frac{d^4k^\prime}{\pi^2i}\hspace*{0.1cm}
{\rm Tr}\bigg[\Gamma_1^\prime S_q\biggl(\frac{k^\prime-k}{2}\biggr)
\Gamma_2^\prime S_q\biggl(\frac{k^\prime+k}{2}\biggr)\biggr] \\[5mm]
&\times&\exp\biggl(\frac{18k^2+6k^{\prime 2}}{\Lambda^2_{B_Q}}\biggr)
\bar u(v^\prime)\Gamma_1 S_{v^\prime}(k,\bar\Lambda)
\Gamma S_v(k,\bar\Lambda)\Gamma_2u(v)\nonumber
\end{eqnarray}
\vspace*{0.3cm}
$\bullet$ $c\to s$ transition
\begin{eqnarray}\label{CS-trans}
\bar u(p^\prime)M_\Gamma(p,v^\prime)u(v)&=&
g_{B_s}g_{B_c}\hspace*{-0.1cm}\int\hspace*{-0.1cm}\frac{d^4k}{\pi^2i}
\hspace*{-0.1cm}\int\hspace*{-0.1cm}\frac{d^4k^\prime}{\pi^2i}\hspace*{0.1cm}
{\rm Tr}\bigg[\Gamma_1^\prime S_q\biggl(\frac{k^\prime-k}{2}\biggr)
\Gamma_2^\prime S_q\biggl(\frac{k^\prime+k}{2}\biggr)\biggr]\\[5mm]
&\times&\exp\biggl(\frac{9k^2+3k^{\prime 2}}{\Lambda^2_{B_Q}}\biggr)
\exp\biggl(\frac{9(k+\alpha p^\prime)^2+3k^{\prime 2}}{\Lambda^2_{B_q}}\biggr)
\nonumber\\[5mm]
&\times&\bar u(p^\prime)\Gamma_1 S_s(k+p^\prime)
\Gamma S_v(k,\bar\Lambda)\Gamma_2u(v)
\nonumber\\[5mm]
& &\alpha=\frac{2m}{2m+m_s}\nonumber
\end{eqnarray}

\noindent Here ${\rm Tr}[...]$ corresponds to the light quark loop
obtained after a standard transformations which involves
the charge conjugation matrix $C$

\begin{eqnarray}
(C\Gamma_1^\prime)^{\alpha\mu}S^{\mu\nu}_q\biggl(\frac{k^\prime-k}{2}\biggr)
(\Gamma_2^\prime C)^{\nu\beta}S^{\alpha\beta}_q
\biggl(-\frac{k^\prime+k}{2}\biggr)
={\rm Tr}\bigg[\Gamma_1^\prime S_q\biggl(\frac{k^\prime-k}{2}\biggr)
\Gamma_2^\prime S_q\biggl(\frac{k^\prime+k}{2}\biggr)\biggr]
\nonumber
\end{eqnarray}
\noindent Calculational details of the matrix elements
(\ref{BC-trans}) and (\ref{CS-trans}) are given in Appendix A.

\begin{figure}[htbp]
\unitlength=1.00mm
\special{em:linewidth 0.4pt}
\linethickness{0.4pt}
\begin{center}
\begin{picture}(127.00,90.00)
\put(70.00,23.00){\oval(48.00,16.00)[]}
\put(45.00,23.00){\circle*{5.20}}
\put(95.00,23.00){\circle*{5.00}}
\put(95.00,23.00){\line(1,0){25.00}}
\put(120.00,24.00){\line(-1,0){25.00}}
\put(95.00,22.00){\line(1,0){25.00}}
\put(45.00,22.00){\line(-1,0){25.00}}
\put(20.00,23.00){\line(1,0){25.00}}
\put(45.00,24.00){\line(-1,0){25.00}}
\put(44.00,24.00){\line(3,5){26.00}}
\put(96.00,24.00){\line(-3,5){26.00}}
\put(70.00,67.00){\circle*{2.00}}
\put(70.00,67.00){\line(0,1){3.00}}
\put(70.00,71.00){\line(0,1){3.00}}
\put(70.00,75.00){\line(0,1){3.00}}
\put(70.00,79.00){\line(0,1){3.00}}
\put(70.00,83.00){\line(0,1){3.00}}
\put(70.00,87.00){\line(0,1){3.00}}
\put(75.00,71.00){\line(0,3){15.00}}
\put(75.00,86.00){\line(-1,-3){1.50}}
\put(75.00,86.00){\line(1,-3){1.50}}
\put(86.00,78.00){\makebox(0,0)[cc]{{\bf q=p-p$^\prime$}}}
\put(70.00,96.00){\makebox(0,0)[cc]{{\bf lept. pair}}}
\put(13.00,23.00){\makebox(0,0)[cc]{{\bf B$^\prime$}}}
\put(127.00,23.00){\makebox(0,0)[cc]{{\bf B}}}
\put(110.00,31.00){\makebox(0,0)[cc]{{\bf p}}}
\put(30.00,31.00){\makebox(0,0)[cc]{{\bf p$^\prime$}}}
\put(94.00,47.00){\makebox(0,0)[cc]{{\bf k+p}}}
\put(47.00,47.00){\makebox(0,0)[cc]{{\bf k+p$^\prime$}}}
\put(71.00,36.00){\makebox(0,0)[cc]{{\bf (k$^\prime$-k)/2}}}
\put(71.00,10.00){\makebox(0,0)[cc]{{\bf -(k$^\prime$+k)/2}}}
\put(87.00,22.00){\makebox(0,0)[cc]{$\bf \Gamma^\prime_2 C$}}
\put(53.00,22.00){\makebox(0,0)[cc]{$\bf C \Gamma^\prime_1$}}
\put(97.00,16.00){\makebox(0,0)[cc]{$\bf \Gamma_2$}}
\put(44.00,16.00){\makebox(0,0)[cc]{$\bf \Gamma_1$}}
\put(28.00,23.00){\line(2,1){4.00}}
\put(28.00,23.00){\line(2,-1){4.00}}
\put(108.00,23.00){\line(2,1){4.00}}
\put(108.00,23.00){\line(2,-1){4.00}}
\put(68.00,31.10){\line(4,1){4.00}}
\put(68.00,31.10){\line(4,-1){4.00}}
\put(68.00,15.10){\line(4,1){4.00}}
\put(68.00,15.10){\line(4,-1){4.00}}
\put(56.00,44.00){\line(0,2){4.00}}
\put(56.00,44.00){\line(2,1){4.00}}
\put(82.00,47.30){\line(2,-1){4.00}}
\put(82.00,47.30){\line(0,-2){4.00}}
\put(63.00,69.00){\makebox(0,0)[cc]{{\bf O$_\mu$}}}
\end{picture}
\vspace*{.2cm}
\mbox{{\bf Fig.2}. Semileptonic decay of heavy-light baryon.}
\end{center}
\end{figure}

We now turn to the discussion of matrix elements of heavy-to-heavy
baryonic decays.
In this paper we consider decays of bottom baryons
($\Lambda_b^0$, $\Xi_b^0$, $\Sigma_b^+$ and $\Omega_b^-$) into pseudoscalar
charmed baryons ($\Lambda_c^+$, $\Sigma_c^{++}$ and $\Omega_c^0$) and
pseudovector states ($\Sigma_c^{\star ++}$ and $\Omega_c^{\star +}$).
The matrix elements describing weak transitions between heavy
baryons can be decomposed into a set of relativistic form factors.
In the HQL these form factors are proportional to three universal
functions $\zeta, \xi_1, \xi_2$ of the variable $\omega=v\cdot v^\prime$,
the so-called Isgur-Wise functions \cite{Isgur2,Georgi2}.
The function $\zeta(\omega)$
describes the $b-c$ transitions of $\Lambda$-type baryons.
The functions $\xi_1(\omega)$ and $\xi_2(\omega)$ describe
transitions of $\Omega$-type baryons.

Weak hadronic currents describing the transition of a heavy baryon
$B_b(v)$ with four-velocity $v$ to a heavy baryon
$B_c^{(\star)}(v)$ with $v^\prime$ are written as
\cite{Isgur2}-\cite{Hussain2}

\underline{$\Lambda_b\to\Lambda_c$ Transition}
\begin{eqnarray}
<\Lambda_c(v^\prime)|\bar b \,\Gamma \,c|\Lambda_b(v)>=
\zeta(\omega)\bar u(v^\prime)\Gamma u(v),
\nonumber
\end{eqnarray}

\underline{$\Omega_b\to\Omega_c(\Omega_c^\star)$ Transition}
\begin{eqnarray}
<\Omega_c(v^\prime)\,\,\, \mbox{or}\,\,\, \Omega_c^\star(v^\prime)|
\bar b \,\Gamma \, c|\Lambda_b(v)>=
\bar B^\mu_c(v^\prime) \,\Gamma \,B^\nu_b(v)[-\xi_1(\omega)g_{\mu\nu}
+\xi_2(\omega)v_\mu v^\prime_\nu],
\nonumber
\end{eqnarray}
where the spinor tensor $B^\nu_b(v)$ satisfies the Rarita-Schwinger conditions
$v_\nu B^\nu_b(v)=0$ and $\gamma_\nu B^\nu_b(v)=0$. The spin wavefunctions
are written as
\begin{eqnarray}
B^\mu_Q(v)=\frac{\gamma^\mu+v^\mu}{\sqrt{3}}\gamma^5u_{\Omega_Q}(v)\hspace*{0.5cm}
\mbox{for $\Omega_Q$ states} \hspace*{1cm}\mbox{and}\hspace*{1cm}
B^\mu_Q(v)=u^\mu_{\Omega^\star_Q}(v)\hspace*{0.5cm}
\mbox{for $\Omega^\star_Q$  states}
\nonumber
\end{eqnarray}
where $u_{\Omega_Q}(v)$ is the usual spin $1/2$ spinor and
the spinor $u^\mu_{\Omega^\star_Q}(v)$ is the usual Rarita-Schwinger
spinor. Note that the Ward indentity between the derivative of
the mass operator of heavy-light
baryons and the vertex function (\ref{BC-trans}) with
$\Gamma=\gamma_\mu$ and $v=v^\prime$ ensures the correct normalization
of the functions $\zeta(\omega)$ and $\xi_1(\omega)$ at $\omega=1$.

In the heavy quark limit the matrix element of the transition of
heavy baryon containing a scalar light diquark into light baryons is
described by two relativistic form factors $f_1$ and $f_2$.
For example, the typical hadronic current for $\Lambda_c\to \Lambda^0$
transition is written as
\begin{eqnarray}\label{LCS-martix}
<\Lambda(p^\prime)|\bar sO_\mu c|\Lambda_c(v)>=
\bar u_\Lambda(p^\prime)[f_1(p^\prime\cdot v)+\not \! v
f_2(p^\prime\cdot v)]O_\mu u_{\Lambda_c}(v)
\nonumber
\end{eqnarray}

\vspace*{1cm}
\section{Results}
In this section we give numerical results on the observables of
semileptonic decays of bottom and charm baryons:
the baryonic Isgur-Wise functions, decay rates and asymmetry parameters
in the two-cascade decays
$\Lambda_b\to\Lambda_c[\to \Lambda_s\pi]+W[\to\ell\nu_\ell]$
and $\Lambda_c\to\Lambda_s[\to p\pi]+W[\to\ell\nu_\ell]$.
Our model contains a number of parameters. The cutoff parameter
$\Lambda_{B_q}$
and the light quark mass  $m_q$ are taken from a fit to proton and neutron
data \cite{PSI}. The cutoff parameter $\Lambda_{B_Q}$ relevant for heavy-light
baryons, the binding energy $\bar\Lambda=M_{B_Q}-m_Q$ and the strange quark
mass $m_s$ are fixed by comparison with the experimentally measured decay
$\Lambda^+_c\to\Lambda^0+e^+ +\nu_e$.
We have checked that the Isgur-Wise functions $\xi_1$ and $\xi_2$
satisfy the model-independent Bjorken-Xu inequalities \cite{Xu}.
We give a detailed description of the $\Lambda^+_c\to\Lambda^0+e^+ +\nu_e$
decay, which was recently measured by
CLEO Collaboration \cite{CLEO}. In what follows we will use
the following values for the CKM matrix elements: $|V_{bc}|$=0.04,
$|V_{cs}|$=0.975.

\subsection{Baryonic Isgur-Wise Functions}

In sec. 2 we have introduced heavy-light baryonic currents. We present
a full list of possible currents (without derivatives) with the
quantum numbers of baryons $J^P={\frac{1}{2}}^+$ and $J^P={\frac{3}{2}}^+$.
For simplicity we restrict ourselves
to only one variant of the three-quark currents
for each kind of heavy-light baryon:
{\it pseudoscalar current} (\ref{pseud-h}) for $\Lambda_Q$-type baryons and
{\it vector currents} (\ref{vect1-h},\ref{vect2-h}) for
$\Omega_Q$-type baryons. A justification of this procedure may be taken from
the QCD sum rule analysis of \cite{Groote} where it was found that,
using the  {\it axial current} for $\Lambda_Q$ baryons and
{\it tensor currents} for $\Omega^{(\star)}_Q$
baryons, one obtains results which are not very different from the ones
with the {\it pseudoscalar current} and the {\it vector currents}.
The direct calculation of the IW-functions with currents (\ref{pseud-h})
and (\ref{vect1-h},\ref{vect2-h}) gives the following results
\begin{eqnarray}
& &\zeta(\omega)=\frac{F_0(\omega)}{F_0(1)}, \;\;
\xi_1(\omega)=\frac{F_1(\omega)}{F_1(1)}, \;\;
\xi_2(\omega)=\frac{F_2(\omega)}{F_1(1)}\label{expres}\\
& &\nonumber\\
& &F_{\rm I}(\omega)=
\int\limits_0^\infty dxx\int\limits_0^\infty \frac{dyy}{(y+1)^2}
\int\limits_0^1 d\phi\int\limits_0^1 d\theta \,\, R_I(\omega) \,\,
\exp\biggl[-6S(\beta)(4\mu_q^2-\bar\lambda^2)\biggr]
\nonumber \\
& &\nonumber\\
&\times&\exp\biggl[-12x^2S(\beta)\phi(1-\phi)(\omega-1)
-6S(\beta)(x-\bar\lambda)^2-24\mu_q^2(1-2\theta)^2\frac{y^2}{1+y}\biggr]
\nonumber
\end{eqnarray}
\noindent where

\begin{eqnarray}
R_0(\omega)&=&\mu_q^2+\frac{1}{6S(\beta)(1+y)}+
\frac{x^2\beta}{4(1+y)^2}(1+2\phi(1-\phi)(\omega-1))\nonumber \\
& &\nonumber\\
R_1(\omega)&=&\mu_q^2+\frac{1}{12S(\beta)(1+y)}+
\frac{x^2\beta}{4(1+y)^2}(1+2\phi(1-\phi)(\omega-1))\nonumber \\
& &\nonumber\\
R_2(\omega)&=&\frac{x^2\beta}{2(1+y)^2}
\phi(1-\phi)=\frac{R_1(\omega)-R_1(1)}{\omega-1}\nonumber \\
& &\nonumber\\
\beta&=&1+2y+4y^2\theta(1-\theta), \,\,\,\,\,\,
S(\beta)=\frac{2}{3}+\frac{\beta}{3(1+y)},
\,\,\,\,\,\, \mu_q=\frac{m_q}{\Lambda_Q},\,\,\,\,
\bar\lambda=\frac{\bar\Lambda}{\Lambda_Q} \,\,\,\,\,
\nonumber
\end{eqnarray}
\noindent All mass-dimension variables are scaled by the parameter
$\Lambda_Q$. Hence the IW-functions depend only on two parameters
$\mu_q$ and $\bar\lambda$. We reiterate  that the functions $\zeta$ and $\xi_1$
are normalized to one at zero recoil due to the existence of a Ward
identity relating the vertex function with the derivative of the heavy-light
baryon mass operator as discussed after Eq.(\ref{ff}).
Contrary to this the normalization of the $\xi_2$-function is model-dependent.
In our model the value $\xi_2(1)$ satisfies the inequality
$0 < \xi_2(1) < 1/2$ and depends on the choice of the
parameters $\mu_q$ and $\bar\lambda$.

It is easy to show that the baryonic IW-functions can be rewritten
in the form

\begin{eqnarray}
\hspace*{-.6cm}& &\zeta(\omega)=
\frac{\sum\limits_{N=0}^\infty C_N^\zeta \bar\lambda^N \Phi_N(\omega)}
{\sum\limits_{N=0}^\infty C_N^\zeta\bar\lambda^N}\leq
\Phi_0(\omega)=\frac{\ln(\omega+\sqrt{\omega^2-1})}{\sqrt{\omega^2-1}},
\nonumber\\[5mm]
\hspace*{-.6cm}& &\xi_1(\omega)=
\frac{\sum\limits_{N=0}^\infty C_N^{\xi_1} \bar\lambda^N \Phi_N(\omega)}
{\sum\limits_{N=0}^\infty C_N^{\xi_1}\bar\lambda^N}\leq
\Phi_0(\omega)=\frac{\ln(\omega+\sqrt{\omega^2-1})}{\sqrt{\omega^2-1}},
\nonumber\\[5mm]
\hspace*{-.6cm}& &\xi_2(\omega)=\frac{\sum\limits_{N=0}^\infty C_N^{\xi_2}
\bar\lambda^N(\Phi_N(\omega)-\Phi_{N+1}(\omega))}{(\omega - 1)
\sum\limits_{N=0}^\infty C_N^{\xi_1}\bar\lambda^N} <
\frac{\Phi_0(\omega)-\Phi_1(\omega)}{\omega - 1}=\frac{1}{\omega^2 - 1}
\biggl(\frac{\omega\ln(\omega+\sqrt{\omega^2-1})}{\sqrt{\omega^2-1}}-1\biggr)
\nonumber
\end{eqnarray}
Here
\begin{eqnarray}
\Phi_N(\omega)&=&\int\limits_0^1
\frac{d\phi}{[1+2(w-1)\phi(1-\phi)]^{N/2+1}} \leq \Phi_0(\omega)\,\,\,
\mbox{for}\,\,\,\forall N\geq 0,
\nonumber\\[5mm]
C_N^F&=&\frac{(2\sqrt{6})^N\Gamma(N/2+1)}{12\Gamma(N)}
\int\limits_0^1 d\theta\int\limits_0^\infty dyy \frac{S^{N/2-1}(\beta)}
{(1+y)^2}\exp(-24\mu_q^2y)\Delta_F > 0, \,\,\,F=\zeta, \xi_1, \xi_2
\nonumber\\[5mm]
\Delta_\zeta&=&\mu_q^2+\frac{1}{6S(\beta)(1+y)}+
\biggl(\frac{N}{2}+1\biggr)\frac{\beta}{24S(\beta)(1+y)^2}\nonumber\\[5mm]
\Delta_{\xi_1}&=&\mu_q^2+\frac{1}{12S(\beta)(1+y)}+
\biggl(\frac{N}{2}+1\biggr)\frac{\beta}{24S(\beta)(1+y)^2}\nonumber\\[5mm]
\Delta_{\xi_2}&=&
\biggl(\frac{N}{2}+1\biggr)\frac{\beta}{24S(\beta)(1+y)^2}\nonumber\\[5mm]
\Gamma(N)&=&\int\limits_0^\infty dt t^{N-1} \exp(-t) \,\,\,\mbox{is the
$\gamma$ - function}\nonumber
\end{eqnarray}

Note that $\zeta(\omega)$ and $\xi_1(\omega)$ become largest
when $\bar\lambda=0$.
\begin{eqnarray}\label{bounds}
\zeta(\omega)\equiv\xi_1(\omega)\equiv\Phi_0(\omega)
\end{eqnarray}
\noindent An increase of $\bar\lambda$ leads to
a suppression of the IW-functions in the physical
kinematical region of the variable $\omega$, i.e. in the region
\begin{eqnarray}\label{region}
1\leq\omega\leq\omega_{max}=\frac{M^2_{B_Q}+M^2_{B_Q^\prime}}
{2M_{B_Q}M_{B_Q^\prime}}
\end{eqnarray}

The radii of the form factors $\zeta$ and $\xi_1$ are defined as
\begin{eqnarray}\label{radius}
F(\omega)=1-\rho^2_F(\omega-1)+...,  \hspace*{1cm}F=\zeta, \xi_1
\end{eqnarray}

It is easy to show that $\rho^2_\zeta$ and $\rho^2_{\xi_1}$ have
the lower bound
\begin{eqnarray}\label{rad}
& &\rho^2_\zeta = \frac{1}{3}+2 \,\,\bar\lambda \,\,
\frac{I(2,2)}{I(1,2)} \geq \frac{1}{3},\hspace*{1cm}
\rho^2_{\xi_1} = \frac{1}{3}+2 \,\,\bar\lambda \,\,
\frac{I(2,1)}{I(1,1)} \geq \frac{1}{3}
\end{eqnarray}
since the integral
\begin{eqnarray}
I(M,N)&=&\int\limits_0^1 d\theta \int\limits_0^\infty dx
\int\limits_0^\infty \frac{dyy}{(1+y)^2} S(\beta) x^M
\biggl[\mu_q^2+\frac{N}{12S(\beta)(1+y)}+\frac{x^2\beta}{4(1+y)^2}\biggr]
\nonumber\\[5mm]
&\times&\exp\biggl[-6S(\beta)x(x-2\bar\lambda)-24\mu^2_qy\biggr]
\nonumber
\end{eqnarray}
is always positive.

As was shown in \cite{Xu}, the IW-functions $\xi_1$ and $\xi_2$ must
respect the two model-independent Bjorken-Xu inequalities.
The first inequality
\begin{eqnarray}\label{ineq1}
1\geq B(\omega)=\frac{2+\omega^2}{3}\xi_1^2(\omega)+
\frac{(\omega^2-1)^2}{3}\xi_2^2(\omega)
+\frac{2}{3}(\omega-\omega^3)\xi_1(\omega)\xi_2(\omega)
\end{eqnarray}
\noindent
is derived from the Bjorken sum rule for semileptonic $\Omega_b$ decays to
the ground state and to low-lying negative-parity excited charmed baryon
states in the HQL. The inequality (\ref{ineq1}) implies a second
inequality, namely a model-independent restriction of the slope (radius) of the
form factor $\xi_1(\omega)$
\begin{eqnarray}\label{ineq2}
\rho^2_{\xi_1}\geq \frac{1}{3}-\frac{2}{3}\xi_2(1)
\end{eqnarray}
\noindent Let us check whether our IW-functions $\xi_1$ and $\xi_2$
respect these inequalities.
First, the inequality (\ref{ineq2}) for the slope of the $\xi_1$-function
can be seen to be satisfied because from Eq.(\ref{rad}) one has
$\rho^2_{\xi_1} \geq 1/3$ and further $\xi_2(1) > 0$ from Eq.(\ref{expres}).

To check the inequality (\ref{ineq1}) we rewrite it in the form
\begin{eqnarray}\label{ineq3}
1\geq B(\omega)=\frac{2}{3}\xi_1^2(\omega)+
\frac{1}{3}(\omega\xi_1(\omega)-\xi_2(\omega)(\omega^2-1))^2
\end{eqnarray}
\noindent One can show that the combination $\omega\xi_1(\omega)-
\xi_2(\omega)(\omega^2-1)$ satisfies the following condition
$\xi_1(\omega)\leq\omega\xi_1(\omega)-\xi_2(\omega)(\omega^2-1)
\leq\omega\xi_1(\omega)$.
Hence,
\begin{eqnarray}\label{ineq4}
\xi_1^2(\omega) \leq B(\omega) \leq \frac{2+\omega^2}{3}\xi_1^2(\omega)
\end{eqnarray}
\noindent From the inequalities (\ref{ineq3}) and (\ref{ineq4})
one finds an upper limit for the function $\xi_1(\omega)$:
\begin{eqnarray}\label{upper}
\xi_1(\omega)\leq\sqrt{\frac{3}{2+\omega^2}}
\end{eqnarray}

The results for the IW-functions $\zeta(\omega)$ and $\xi_1(\omega)$ are
plotted in Fig.3-7 in the physical region $1\leq \omega \leq \omega_{max}$.
The function $\xi_1(\omega)$ is shown for the two cases:
a) decay of $\Sigma_b$-baryon and b) decay of $\Omega_b$-baryon.
In Figs.3 and 4 we demonstrate the sensitivity of the $\zeta$-function
on the choice of the parameters $\bar\Lambda$ and $\Lambda_Q$ when
one is varied and the other one is fixed.
Below (in sec.4.2) we will show that the best description
of the experimental data for $\Lambda^+_c\to\Lambda^0+e^+ +\nu_e$ decay
is obtained with the choice of parameters $\Lambda_Q$=2.5 GeV,
$\bar\Lambda$=710 MeV and $m_s$=570 MeV. In Fig.3 the $\zeta(\omega)$
function is shown for $\bar\Lambda$ values between 600 MeV to 800 MeV,
where the parameter $\Lambda_Q$
is assumed to be 2.5 GeV. It is seen that an increase of $\bar\Lambda$
leads to a suppression of the baryonic IW-function $\zeta$. In Fig.4 the
dependence of $\zeta$ on the value $\Lambda_Q$ is plotted for
$\bar\Lambda$=710 MeV. One can see that a decrease of $\Lambda_Q$ leads
to a suppression of $\zeta(\omega)$.
In  Fig.5 we give the {\it best fit} for the IW-function $\zeta$
($\Lambda_Q$=2.5 GeV, $\bar\Lambda$=710 MeV).
For comparison the results of other phenomenological
approaches are shown too where we compare with results obtained from QCD sum
rules \cite{Grozin}, IMF models \cite{Kroll,Koerner2}, MIT bag model
\cite{Zalewski}, a simple quark model (SQM) \cite{Mark1} and
the dipole formula \cite{Koerner2}. Our result is close to the QCD sum rule
result \cite{Grozin}. For quick reference we want to remark that
in the physical region our function $\zeta$ can be well approximated
by the formula
\begin{eqnarray}\label{approx}
\zeta(\omega)\approx \biggl[\frac{2}{1+\omega}\biggr]^{1.7+1/\omega}
\end{eqnarray}
In Fig.6 and 7 we analyse the $\omega$-dependence of the $\xi_1$ form factor.
We exhibit the dependence of $\xi_1(\omega)$
on the choice of $\bar\Lambda$ for $\Sigma_b$ baryon
decays (Fig.6) and for $\Omega_b$ baryon
decays (Fig.7). For both cases $\Lambda_Q$ is put equal to 2.5 GeV. In the
analysis of the $\Omega_b$ form factor we use $m_s=570$ MeV.
We also present results on the upper limit (22) for the function
$\xi_1(\omega)$. In Fig.7 we also compare to a simple quark model
calculation of \cite{Mark2}. We want to emphasize that for both cases,
$\Sigma_b$ and $\Omega_b$ baryon decays, our $\zeta_1$ does
not exceed the upper limit (\ref{upper})
except in a narrow region of very small (unphysical) values of
$\bar\Lambda$: $\bar\Lambda\leq $ 60 MeV.
Thus we conclude that the Bjorken-Xu inequality is respected by our model.

The results for the charge radii are listed in Tables 3-6 for various sets of
the adjustable parameters.
For comparison we quote the results for the charge radii predicted
by other phenomenological approaches:
$\rho^2_\zeta=3.04$ (IMF model) \cite{Koerner2},  $\rho^2_\zeta=1.78$
(dipole formula) \cite{Koerner2}, $\rho^2_\zeta=2.28$ (MIT bag model)
\cite{Zalewski}, $\rho^2_{\zeta}=1$ and $\rho^2_{\xi_1}=1.02\div 1.18$
(simple quark model) \cite{Mark1,Mark2}, $\rho^2_{\zeta}=0.55\pm 0.15$
(QCD Sum Rules) \cite{Dai}.

\vspace*{1.5cm}
\begin{center}
{\bf Table 3}. The Charge Radius $\rho^2_{\zeta}$
of $\Lambda_b$ baryon at $\Lambda_Q$=2.5 GeV.\\
\end{center}
\def\arraystretch{2.}
\begin{center}
\begin{tabular}{|c|c|c|c|c|c|c|c|c|c|c|c|c|}
\hline\hline
$\bar\Lambda$ (MeV) & 600  & 625 & 650  & 675 & 700  & 710  & 725
                    & 750  & 675 & 690  & 800 \\
\hline
$\rho^2_{\zeta}$    & 1.04 & 1.09 & 1.12 & 1.22 & 1.30 & 1.33 & 1.38
                    & 1.47 & 1.59 & 1.68 & 1.76 \\
\hline\hline
\end{tabular}
\end{center}

\vspace*{1cm}
\begin{center}
{\bf Table 4.} The Charge Radius $\rho^2_{\zeta}$
of $\Lambda_b$ baryon at $\bar\Lambda$=710 MeV.\\
\end{center}
\def\arraystretch{2.}
\begin{center}
\begin{tabular}{|c|c|c|c|c|c|c|c|c|c|c|c|c|}
\hline\hline
$\Lambda_Q$ (GeV) & 1.25 & 1.3 & 1.4 & 1.5 & 1.6 & 1.7 & 1.8
                  & 1.9  & 2.0 & 2.1 & 2.3 & 2.5 \\
\hline
$\rho^2_{\zeta}$  & 2.93 & 2.82 & 2.63 & 2.47 & 2.31 & 2.15 & 2.0
                  & 1.87 & 1.75 & 1.64 & 1.46 & 1.33 \\
\hline\hline
\end{tabular}
\end{center}

\vspace*{1cm}
\begin{center}
{\bf Table 5.} The Charge Radius $\rho^2_{\xi_1}$
of $\Sigma_b$ baryon at $\Lambda_Q$=2.5 GeV.\\
\end{center}
\def\arraystretch{2.}
\begin{center}
\begin{tabular}{|c|c|c|c|c|c|c|c|c|c|c|c|c|}
\hline\hline
$\bar\Lambda$ (MeV) & 600  & 625 & 650  & 675 & 700  & 710  & 725
                    & 750  & 675 & 690  & 800 \\
\hline
$\rho^2_{\xi_1}$    & 1.05 & 1.09 & 1.12 & 1.22 & 1.32 & 1.35
                    & 1.38 & 1.50 & 1.59 & 1.68 & 1.80 \\
\hline\hline
\end{tabular}
\end{center}

\vspace*{1cm}
\begin{center}
{\bf Table 6.} The Charge Radius $\rho^2_{\xi_1}$
of $\Omega_b$ baryon at $\Lambda_Q$=2.5 GeV.\\
\end{center}
\begin{center}
\def\arraystretch{2.}
\begin{tabular}{|c|c|c|c|c|c|c|c|c|c|c|c|c|c|}
\hline\hline
$\bar\Lambda_{\{ss\}}$ (MeV) & 800  & 850  & 875  & 900 & 925 & 950 & 975 & 1000
                         & 1025 & 1050 & 1075 & 1100 \\
\hline
$\rho^2_{\xi_1}$     & 1.44& 1.58& 1.66 & 1.74& 1.82& 1.92& 2.02 & 2.12
                     & 2.25& 2.39 & 2.56& 2.79\\
\hline\hline
\end{tabular}
\end{center}

\vspace*{1.5cm}
\subsection{Rates, Distributions and Asymmetry Parameters in $b\to c$
Baryonic Decays}

In this section we present on numerical results for rates,
distributions and asymmetry
parameters in the $b\to c$ flavor changing baryon decays.
The standard expressions for observables
of semileptonic decays of bottom baryons (decay rates, differential
distributions, leptonic spectra and asymmetry parameters) have
simple forms when expressed in  terms of helicity  amplitudes
$H_{\lambda_f\lambda_W}$ \cite{Kramer,DESY}, where $\lambda_f$
is helicity of the final state baryon and $\lambda_W$ is the helicity of
the off mass-shell W-boson. The HQL helicity amplitudes
describing transitions of bottom baryon into charm ones are
expressed through IW-functions in the following way:

\[H_{\pm\frac{1}{2}\pm 1}=
-2\sqrt{M_iM_f}(\sqrt{\omega-1}\mp\sqrt{\omega+1})\times\left\{
\begin{array}{ll}
\zeta(\omega) & \Lambda_b\to\Lambda_c \mbox{decay} \\
\frac{1}{3}\xi_T(\omega)
& \Omega_b\to\Omega_c \mbox{decay} \\
\pm\frac{\sqrt{2}}{3}\xi_T(\omega)
& \Omega_b\to\Omega_c^\star \mbox{decay} \\
\end{array}
\right.  \]

\vfill
\[H_{\pm\frac{1}{2}0}=
\frac{1}{\sqrt{\omega_{max}-\omega}}\times\left\{
\begin{array}{ll}
\zeta(\omega)[M_+\sqrt{\omega-1}\mp M_-\sqrt{\omega+1}]
& \Lambda_b\to\Lambda_c \mbox{decay} \\
\frac{1}{3}[M_+\sqrt{\omega-1}\xi_{L_+}(\omega)\mp M_-
\xi_{L_-}(\omega)\sqrt{\omega+1}]
& \Omega_b\to\Omega_c \mbox{decay} \\
\frac{\sqrt{2}}{3}
[M_+\sqrt{\omega-1}\xi_{L^\star_+}(\omega)\mp M_-\xi_{L^\star_-}(\omega)
\sqrt{\omega+1}] & \Omega_b\to\Omega_c^\star \mbox{decay} \\
\end{array}
\right.  \]

\vfill
\begin{eqnarray}
H_{\pm\frac{3}{2}\pm 1}=\mp 2\xi_1(\omega)
\sqrt{\frac{2}{3}M_iM_f}[\sqrt{\omega-1}\mp \sqrt{\omega+1}]
\hspace*{1cm}\Omega_b\to\Omega_c^\star \mbox{decay}
\nonumber
\end{eqnarray}

\noindent where
\begin{eqnarray}
& &M_\pm=M_i\pm M_f, \hspace*{2.5cm}
\xi_T=\xi_1\omega-\xi_2(\omega^2-1),\nonumber\\[3mm]
& &\xi_{L_\pm}=\xi_1(\omega\pm 2)-\xi_2(\omega^2-1),
\hspace*{0.5cm}\xi_{L^\star_\pm}=\xi_1(\omega\mp 1)-\xi_2(\omega^2-1),
\nonumber \\[3mm]
& &\omega_{max}=\frac{M^2_i+M^2_f}
{2M_iM_f}.\nonumber
\end{eqnarray}

\noindent The decay rates of semileptonic decays are then given by

\begin{eqnarray}
\Gamma=\int\limits_1^{\omega_{max}}d\omega \;\;\frac{d\Gamma}{d\omega},
\;\;\;\frac{d\Gamma}{d\omega}=\frac{d\Gamma_{T_+}}{d\omega}+
\frac{d\Gamma_{T_-}}{d\omega}+\frac{d\Gamma_{L_+}}{d\omega}+
\frac{d\Gamma_{L_-}}{d\omega}
\end{eqnarray}

\noindent where the indices $T$ and $L$ denote partial contributions of
transverse $(\lambda_W=\pm 1)$ and longitudinal $(\lambda_W=0)$
components of the current transitions.
Partial differential distributions are given by
\[\frac{d\Gamma_{T_\pm}}{d\omega}=\kappa_\omega\times\left\{
\begin{array}{ll}
|H_{\pm\frac{1}{2}\pm 1}|^2 & \mbox{for} {\frac{1}{2}}^+\to{\frac{1}{2}}^+
\mbox{transition} \\
|H_{\pm\frac{1}{2}\pm 1}|^2+|H_{\pm\frac{3}{2}\pm 1}|^2 &
\mbox{for} {\frac{1}{2}}^+\to{\frac{3}{2}}^+ \mbox{transition}\\
\end{array}
\right.  \]

\begin{eqnarray}
\hspace*{1.5cm}
\frac{d\Gamma_{L_\pm}}{d\omega}=\kappa_\omega |H_{\pm\frac{1}{2}0}|^2,
\hspace*{1.5cm}\kappa_\omega=\frac{G_F^2}{(2\pi)^3}|V_{bc}|^2
\frac{M^3_f}{6}(\omega_{max}-\omega)\sqrt{\omega^2-1}\nonumber
\end{eqnarray}

Tables 7-11 list our predictions for the semileptonic rates of beauty baryons.
In Table 7 we present the results for total and partial rates
for various $b\to c$ decay modes. The adjustable parameters are
chosen as $m_s=$570 MeV, $\Lambda_Q=$2.5 GeV, $\bar\Lambda=$710 MeV,
$\bar\Lambda_s=$850 MeV and $\bar\Lambda_{\{ss\}}=$1000 MeV.
In Table 8 we compare our results for total rates with
the predictions of other phenomenological approaches: constituent quark
model \cite{DESY}, spectator quark model \cite{Singleton}, nonrelativistic
quark model \cite{Cheng}. The dependence of the total rates on the
parameters $\bar\Lambda$, $\bar\Lambda_s$ and $\bar\Lambda_{\{ss\}}$
are shown in Tables 9-11.

\vspace*{.5cm}
\begin{center}
{\bf Table 7.} Decay Rates of Bottom Baryons (in $10^{10}$ sec$^{-1}$)
for $|V_{bc}|$=0.04\\
\end{center}
\begin{center}
\def\arraystretch{2.}
\begin{tabular}{|c|c|c|c|c|c|c|c|}
\hline\hline
Process & $\Gamma_{total}$ & $\Gamma_T$ & $\Gamma_{T_+}$ & $\Gamma_{T_-}$ &
$\Gamma_L$ & $\Gamma_{L_+}$ & $\Gamma_{L_-}$ \\
\hline
$\Lambda^0_b\to \Lambda^+_c e^-\bar{\nu}_e$
& 5.39 & 2.07 & 0.53 & 1.54 & 3.32 & 0.11 & 3.21\\
\hline
$\Xi^0_b\to \Xi^+_c e^-\bar{\nu}_e$
& 5.27 &  2.02 & 0.54 & 1.48 & 3.25 & 0.11 & 3.14\\
\hline
$\Sigma^+_b\to \Sigma^{++}_c e^-\bar{\nu}_e$
& 2.23 & 0.33 & 0.08 & 0.25 & 1.90 & 1.49 & 0.41\\
\hline
$\Omega^-_b\to \Omega^{0}_c e^-\bar{\nu}_e$
& 1.87 & 0.29 & 0.08 & 0.21 & 1.58 & 1.26 & 0.32\\
\hline
$\Sigma^+_b\to \Sigma^{*++}_c e^-\bar{\nu}_e$
& 4.56 & 2.07 & 0.54 & 1.53 & 2.49 & 1.09 & 1.40\\
\hline
$\Omega^-_b\to \Omega^{*0}_c e^-\bar{\nu}_e$
& 4.01 & 1.89 & 0.53 & 1.36 & 2.12 & 0.95 & 1.17\\
\hline\hline
\end{tabular}
\end{center}

\vspace*{1cm}
\begin{center}
{\bf Table 8.} Model Results for Rates of Bottom Baryons
(in $10^{10}$ sec$^{-1}$)
for $|V_{bc}|$=0.04\\
\end{center}
\begin{center}
\def\arraystretch{2.}
\begin{tabular}{|c|c|c|c|c|} \hline
 Process & Ref.~\cite{Singleton} & Ref.~\cite{Cheng} & Ref.~\cite{DESY}
& Our results\\
\hline\hline
$\Lambda_b^0\to\Lambda_c^+ e^-\bar{\nu}_e$  & 5.9 & 5.1 & 5.14 & 5.39
\\
\hline
$\Xi_b^0\to\Xi_c^+ e^-\bar{\nu}_e$  & 7.2 & 5.3 & 5.21& 5.27
\\
\hline
$\Sigma_b^+\to\Sigma_c^{++} e^-\bar{\nu}_e$
& 4.3 & & & 2.23  \\
\hline
$\Sigma_b^{+}\to\Sigma_c^{\star ++} e^-\bar{\nu}_e$
 & & & &4.56 \\
\hline
$\Omega_b^-\to\Omega_c^0 e^-\bar{\nu}_e$
& 5.4 & 2.3 & 1.52 & 1.87\\
\hline
$\Omega_b^-\to\Omega_c^{\star 0} e^-\bar{\nu}_e$
 & & & 3.41 & 4.01 \\
\hline\hline
\end{tabular}
\end{center}

\vspace*{1cm}
\begin{center}
{\bf Table 9.} Dependence of Rates on $\bar\Lambda$
for $|V_{bc}|$=0.04\\
\end{center}
\def\arraystretch{2.}
\begin{center}
\begin{tabular}{|c|c|c|c|c|c|}
\hline\hline
 Process & \multicolumn{5}{|c|} {$\bar\Lambda$ (MeV)} \\
\cline{2-6}
 & 600 & 650 & 710 & 750 & 800\\
\hline\hline
$\Lambda^0_b\to \Lambda^+_c e^-\bar{\nu}_e$  & 6.10  & 5.83 & 5.39 & 5.19 & 4.74\\
\hline
$\Sigma^+_b\to \Sigma^{++}_c e^-\bar{\nu}_e$ & 2.51 & 2.39 & 2.23 & 2.11 & 1.92 \\
\hline
$\Sigma^+_b\to \Sigma^{*++}_c e^-\bar{\nu}_e$
& 4.99 & 4.81 & 4.56 & 4.35 & 4.03 \\
\hline\hline
\end{tabular}
\end{center}

\newpage
\vspace*{1cm}
\begin{center}
{\bf Table 10.} Dependence of Rates on $\bar\Lambda_s$
for $|V_{bc}|$=0.04\\
\end{center}
\def\arraystretch{2.}
\begin{center}
\begin{tabular}{|c|c|c|c|c|}
\hline\hline
 Process & \multicolumn{4}{|c|} {$\bar\Lambda_s$ (MeV)} \\
\cline{2-5}
 & 760 & 800 & 850 & 900\\
\hline\hline
$\Xi^0_b\to \Xi^+_c e^-\bar{\nu}_e$ & 5.81  & 5.58 & 5.27 & 4.93\\
\hline\hline
\end{tabular}
\end{center}

\vspace*{1cm}
\begin{center}
{\bf Table 11.} Dependence of Rates on $\bar\Lambda_{\{ss\}}$
for $|V_{bc}|$=0.04\\
\end{center}
\def\arraystretch{2.}
\begin{center}
\begin{tabular}{|c|c|c|c|c|c|c|}
\hline\hline
 Process  & \multicolumn{5}{|c|}
{$\bar\Lambda_{\{ss\}}$ (MeV)} \\
\cline{2-6}
 & 900 & 950 & 1000 & 1050 & 1100\\
\hline\hline
$\Omega^-_b\to \Omega^{0}_c e^-\bar{\nu}_e$
& 2.09 & 1.98 & 1.87 & 1.72 & 1.54\\
\hline
$\Omega^-_b\to \Omega^{*0}_c e^-\bar{\nu}_e$
& 4.44 & 4.23 & 4.01 & 3.75 & 3.43\\
\hline\hline
\end{tabular}
\end{center}

\vspace*{.5cm}
The differential distributions for
$\Lambda_b^0\to\Lambda_c^+ e^-\bar\nu$ decay are plotted in Fig.8.

Leptonic spectra $d\Gamma/dE_\ell$ are calculated according to the sum
\begin{eqnarray}
\frac{d\Gamma}{dE_\ell}=\frac{d\Gamma_{T_+}}{dE_\ell}
+\frac{d\Gamma_{T_-}}{dE_\ell}+\frac{d\Gamma_{L_+}}{dE_\ell}
+\frac{d\Gamma_{L_-}}{dE_\ell}
\end{eqnarray}
\noindent Expressions for partial leptonic spectra are given by
\begin{eqnarray}
& &\frac{d\Gamma_{T_\pm}}{dE_\ell}=\hspace*{-0.5cm}\int\limits_{\omega_{min}
(E_\ell)}^{\omega_{max}}\hspace*{-0.5cm}d\omega\hspace*{0.1cm}
\kappa_E(1\pm \cos\Theta)^2
|H_{\pm\frac{1}{2}\pm 1}|^2\;\;\;
\frac{d\Gamma_{L_\pm}}{dE_\ell}=\hspace*{-0.5cm}\int\limits_{\omega_{min}
(E_\ell)}^{\omega_{max}}\hspace*{-0.5cm}d\omega\hspace*{0.1cm}
\kappa_E(1-\cos^2\Theta)^2
|H_{\pm\frac{1}{2}0}|^2, \nonumber\\[5mm]
& &\kappa_E=\frac{G_F^2}{(2\pi)^3}|V_{bc}|^2
\frac{M^2_{\Lambda_c}}{8}(\omega_{max}-\omega),
\;\;\;
\cos\Theta=\frac{E_\ell^{max}-2E_\ell+M_{\Lambda_c}(\omega_{max}-\omega)}
{M_{\Lambda_c}\sqrt{\omega^2-1}}, \nonumber \\[5mm]
& &E^{max}_\ell=\frac{M^2_{\Lambda_b}-M^2_{\Lambda_c}}{2M_{\Lambda_b}},
\;\;\;
\omega_{min}(E_\ell)=\omega_{max}-2\frac{E_\ell}{M_{\Lambda_c}}
\frac{E^{max}_\ell-E_\ell}{M_{\Lambda_b}-2E_\ell}
\nonumber
\end{eqnarray}
\noindent Our results on leptonic spectra in semileptonic
$\Lambda_b\to\Lambda_c$ transitions are shown in Fig.9.

Finally, we consider the cascade decay
$\Lambda_b\to\Lambda_c[\to \Lambda_s\pi]+W[\to\ell\nu_\ell]$ which is
characterized by a set of asymmetry parameters. The formalism and a detailed
analysis of the asymmetry parameters is presented in  \cite{Kramer,DESY}.
In terms of helicity amplitudes the  asymmetry parameters of nonpolarized
$\Lambda_b$ decays
($\alpha, \alpha^\prime, \alpha^{\prime\prime},
\gamma$) and polarized $\Lambda_b$ decays
($\alpha_P, \gamma_P$) are given by the following expressions
\begin{eqnarray}\label{asymmetry}
& &\alpha=\frac{H^-_T+H^-_L}{H^+_T+H^+_L},\;\;\;
\alpha^\prime=\frac{H^-_T}{H^+_T+2H^+_L},\;\;\;
\alpha^{\prime\prime}=\frac{H^+_T-2H^+_L}{H^+_T+2H^+_L},\;\;\;
\gamma=\frac{2H_\gamma}{H^+_T+H^+_L},\nonumber\\[5mm]
& &\alpha_P=\frac{H^-_T-H^-_L}{H^+_T+H^+_L},
\gamma_P=\frac{2H_{\gamma_P}}{H^+_T+H^+_L},\\[5mm]
& &H^\pm_T=|H_{1/2\;1}|^2\pm|H_{-1/2\;-1}|^2
\;\;\;
H^\pm_L=|H_{1/2\;0}|^2\pm|H_{-1/2\;0}|^2\nonumber\\[5mm]
& &H_\gamma=Re(H_{-1/2\;0}H^*_{1/2\;1}+H_{1/2\;0}H^*_{-1/2\;-1})\;\;\;
H_{\gamma_P}=Re(H_{1/2\;0}H^*_{-1/2\;0})
\nonumber
\end{eqnarray}

\noindent We evaluate the average magnitudes of the asymmetry
parameters ($<\alpha>, <\alpha^\prime>$ etc.) as results of separate
$\omega$ integrations of numerators and denominators. Results for average
magnitudes are given in Table 12. Also the results of paper \cite{Koerner2}
are quoted for comparison.

\vspace*{1cm}
\begin{center}
{\bf Table 12.} Asymmetry parameters of $\Lambda_b$ decay
\end{center}
\def\arraystretch{2.0}
\begin{center}
\begin{tabular}{|c|c|c|c|c|c|c|}
\hline\hline
Model & $\alpha$ & $\alpha^\prime$ & $\alpha^{\prime\prime}$ &
$\gamma$ & $\alpha_P$ & $\gamma_P$ \\
\hline
Our  & -0.76 & -0.12 & -0.53 & 0.56 & 0.39 & -0.16\\
\hline
IMF \cite{Koerner2} & -0.71 & -0.12 & -0.46 & 0.61 & 0.33 & -0.19\\
\hline\hline
\end{tabular}
\end{center}

\vspace*{1.5cm}
\subsection{Heavy-to-Light Baryon Decays}

In this subsection we consider the heavy-to-light semileptonic modes.
In particular the process $\Lambda^+_c\to\Lambda^0+e^++\nu_e$
which was recently
investigated by the CLEO Collaboration \cite{CLEO} is studied in detail.
In the heavy mass limit ($m_C\to\infty$) its transition matrix
element is defined by two form factors $f_1$ and $f_2$ (see,
section 3). Assuming identical dipole forms for the form factors
(as in the model of K\"{o}rner and Kr\"{a}mer \cite{Kramer}),
CLEO found that $R=f_2/f_1=$-0.25$\pm$0.14$\pm$0.08. Our form factors have
different $q^2$ dependences. In other words, the quantity $R=f_2/f_1$
has a $q^2$ dependence in our approach. In Fig.10 we plot the results
for $R$ in the kinematical region $1\leq \omega \leq \omega_{max}$ for
different magnitudes of the $\bar\Lambda$ parameter.

It is seen that larger values of $\bar\Lambda$ lead to an
increase of the ratio $R$. The best fit to the experimental data is achieved
for the following set of parameters: $m_s=$570 MeV, $\Lambda_Q=$2.5 GeV
and $\bar\Lambda=$710 MeV. In this case the $\omega$-dependence of the
form factors $f_1$, $f_2$ and their ratio $R$ are shown in Fig.11.
Particularly, we get $f_1(q^2_{max})$=0.8, $f_2(q^2_{max})$=-0.18,
$R$=\,-0.22 at zero recoil ($\omega$=1 or q$^2$=q$^2_{max}$) and
$f_1(0)$=0.38, $f_2(0)$=-0.06, $R$=-0.16 at maximum recoil
($\omega=\omega_{max}$ or $q^2$=0). Note that our results for
$q^2_{max}$ are close to those of the nonrelativistic quark model \cite{Cheng}:
$f_1(q^2_{max})$=0.75, $f_2(q^2_{max})$=-0.17, $R$=-0.23.

Our result for $R$ agree well with the experimental
data \cite{CLEO} $R=-0.25 \pm 0.14 \pm 0.08$. The predictions for
the decay rate
$\Gamma(\Lambda^+_c\to\Lambda^0e^+\nu_e)$=7.22$\times$ 10$^{10}$ sec$^{-1}$
and for the asymmetry parameter $\alpha_{\Lambda_c}$=-0.812 also coincide
with the experiment: $\Gamma_{exp}$=7.0$\pm$ 2.5 $\times$ 10$^{10}$ sec$^{-1}$
and $\alpha_{\Lambda_c}^{exp}$=-0.82$^{+0.09+0.06}_{-0.06-0.03}$ respectively
as well as with the result of \cite{Cheng}
$\Gamma$=7.1 $\times$ 10$^{10}$ sec$^{-1}.$
Note that the agreement with the experimental
rate measurement crucially depends on the use of the
$\Lambda^0$ three-quark current
in its $SU(3)$-flavor symmetric form (see, Table 1.)
which leads to the presence of the flavor-suppression
factor  $N_{\Lambda_c\Lambda}=1/\sqrt{3}$ for
$\Lambda_c^+\to\Lambda^0 e^+\nu_e$. If the $SU(3)$ symmetric
structure of $\Lambda^0$ hyperon is not taken into account the
predicted rate for $\Lambda_c^+\to\Lambda^0 e^+\nu_e$ becames too large
(see, discussion in ref.~\cite{DESY,Cheng}).

In Table 13 we present our predictions for some modes of
semileptonic heavy-to-light transitions (for $\bar\Lambda_s$=850 MeV,
$\bar\Lambda_{\{ss\}}$=1000 MeV).
Also the results obtained in other approaches are tabulated.
Note that the flavor-suppression
factor for the modes $\Xi_c^0\to\Xi^- e^+\nu_e$,
$\Lambda_b^0\to p e^-\bar\nu_e$ and $\Lambda_c^+\to ne^+\nu_e$
is equal to $1/\sqrt{2}$.

Finally, in Table 14 we give the predictions for the average magnitudes
of the asymmetry parameters for the cascade decay
$\Lambda_c\to\Lambda_s[\to p\pi]+W[\to\ell\nu_\ell]$ which are
expexted to be measured in near future by the COMPASS Collaboration
\cite{COMPASS}.
For comparison, the results of paper \cite{Kramer}
for $R=f_2/f_1=-0.25$ are also given.

\newpage
\begin{center}
{\bf Table 13.} Heavy-to-Light Decay Rates (in 10$^{10}$ s$^{-1}$)
for $|V_{bc}|$=0.04, $|V_{cs}|$=0.975.
\end{center}
\begin{center}
\begin{tabular}{|c|c|c|c|c|c|c|c|} \hline
 Process & Quantity & Ref.\cite{Singleton} & Ref.\cite{Cheng} &
Ref.\cite{Datta} & Ref.\cite{Luo}
& Our & Exp.\cite{PDG} \\
\hline\hline
$\Lambda_c^+\to\Lambda^0 e^+\nu_e$  & $\Gamma$ & 9.8 & 7.1 &
5.36 & 7 & 7.22 & 7.0$\pm$ 2.5 \\
\hline
$\Xi_c^0\to\Xi^- e^+\nu_e$  & $\Gamma$ & 8.5 & 7.4 & & 9.7 & 8.16 & \\
\hline
$\Lambda_b^0\to p e^-\bar\nu_e$ & $\Gamma/|V_{bu}|^2$ & &
& 6.48$\times 10^2$ & & 7.47$\times 10^2$ &\\
\hline
$\Lambda_c^+\to ne^+\nu_e$  & $\Gamma/|V_{cd}|^2$ & & & &
0.17$\times 10^2$ & 0.26$\times 10^2$ & \\
\hline\hline
\end{tabular}
\end{center}

\vspace*{.3cm}

\vspace*{.6cm}
\begin{center}
{\bf Table 14.} Asymmetry parameters of $\Lambda_c$ decay
\end{center}
\def\arraystretch{2.0}
\begin{center}
\begin{tabular}{|c|c|c|c|c|c|c|}
\hline\hline
Model & $\alpha$ & $\alpha^\prime$ & $\alpha^{\prime\prime}$ &
$\gamma$ & $\alpha_P$ & $\gamma_P$ \\
\hline
Our  & -0.81 & -0.13 & -0.56 & 0.50 & 0.40 & -0.15\\
\hline
K\"{o}rner \& Kr\"{a}mer \cite{Kramer}
& -0.82 & -0.13 & -0.56 & 0.47 & 0.39 & -0.14\\
\hline\hline
\end{tabular}
\end{center}

\vspace*{1.5cm}
\section{Conclusion}

We have developed a relativistic model \cite{Aniv1,Aniv2},
\cite{PSI,Manchester} for QCD bound states composed of light quarks
and a heavy quark.
In fact, this model is the Lagrangian formulation of the NJL model with
separable interaction \cite{Goldman,Gross} and its advantage consists in
the possibility of studying baryons as three-quark states as multiquark
and exotic objects. We have used our approach to study the
properties of baryons containing a single heavy quark.
We have calculated the observables of semileptonic decays of
bottom and charm baryons: Isgur-Wise functions, asymmetry parameters,
decay rates and distributions. We obtained analytical
expressions for the baryon IW-functions: $\zeta$ ($\Lambda_b\to\Lambda_c$
transition), $\xi_1$ and $\xi_2$ ($\Omega_b\to\Omega_c^{(\star)}$
transition). We checked the model-independent Bjorken-Xu
inequalities for the $\xi_1$ and $\xi_2$ functions and
their derivatives at the zero recoil point. It is shown that
inequality for the charge radius of $\xi_1$ (see, Eq.~(\ref{ineq2}))
is automatically respected in our model. The inequality (\ref{ineq1}) for
the $\omega$-dependence of $\xi_1$ and $\xi_2$ leads to an upper limit for
$\xi_1$ (see, \ref{upper}) which is respected in our model for
reasonable values of the parameter $\bar\Lambda$.
We have also applied our model to the calculation of heavy-to-light
semileptonic decay processes motivated by the recent experimental
observation of the $\Lambda_c^+\to\Lambda^0 e^+\nu_e$ decay by the CLEO
Collaboration \cite{CLEO}. Our predictions for the form factor ratio,
the decay rate and the asymmetry parameter $\alpha$ are in good agreement
with measured values \cite{CLEO,PDG}. The success in reproducing the correct
experimental rate requires the use of the $\Lambda^0$ three-quark
current in the $SU(3)$-flavor symmetric form (see, Table 1.). Predictions for
other semileptonic heavy-to-light rates are also given. Finally,
we have given predictions for the asymmetry parameters of the cascade decay
$\Lambda_c\to\Lambda_s[\to p\pi]+W[\to\ell\nu_\ell]$.

\vspace*{.5cm}
\section{Acknowledgements}

M.A.I and V.E.L thank Mainz and Wuppertal Universities for the hospitality
where a part of this work was completed.
This work was supported in part by the INTAS Grant 94-739,
the Heisenberg-Landau Program, by the Russian Fund of
Fundamental Research (RFFR) under contract 96-02-17435-a and the
State Committee of the Russian Federation for
Education (project N 95-0-6.3-67,
Grant Center at S.-Petersburg State University).

\newpage

\section{Appendix}

\appendix

\def\appendix{
\vskip 1cm
\par
\setcounter{equation}{0}
\def\theequation{A.\arabic{equation}}
}

\appendix

{\Large\bf A. The Calculation Technique}

\vspace*{0.5cm}
To elucidate the calculation of the matrix elements (\ref{BC-trans}) and
(\ref{CS-trans}) we consider the two generic integrals in a Euclidean space
\begin{eqnarray}
I_1(p^{\prime 2}_E,w_E)&=&
\int\frac{d^4 k_E}{\pi^2}\int\frac{d^4 k^\prime_E}{\pi^2}
\exp\biggl(-\frac{9k^2_E+3k^{\prime 2}_E}{\Lambda_{B_Q}^2}\biggr)
\exp\biggl(-\frac{9(k_E+\alpha p^\prime)^2+3k^{\prime 2}_E}
{\Lambda_{B_q}^2}\biggr)\nonumber\\[4mm]
&\times&\frac{1}{m^2+(k_E+k^\prime_E)^2/4}\,\,
\frac{1}{m^2+(k_E-k^\prime_E)^2/4}\nonumber\\[4mm]
&\times&\frac{1}{m^2_s+(k_E+p^\prime_E)^2}\,\,
\frac{1}{k_Ev_E-\bar\Lambda}; \hspace*{1cm}\alpha=\frac{2m}{2m+m_s}
\end{eqnarray}

\vspace*{0.5cm}
\begin{eqnarray}
I_2(w_E)&=&\int\frac{d^4 k_E}
{\pi^2}\int\frac{d^4 k^\prime_E}{\pi^2}
\exp\biggl(-\frac{18k^2_E+6k^{\prime 2}_E}{\Lambda_Q^2}\biggr)
\nonumber\\[4mm]
&\times&\frac{1}{m^2+(k_E+k^\prime_E)^2/4}\times
\frac{1}{m^2+(k_E-k^\prime_E)^2/4}\nonumber\\[4mm]
&\times&\frac{1}{k_Ev_E-\bar\Lambda}\,\,
\frac{1}{k_Ev^\prime_E-\bar\Lambda}
\end{eqnarray}
\noindent where $\alpha$ is defined in (\ref{CS-trans}).
The final light baryon state carrying the Euclidean momenta
$p^\prime_E$, is on mass-shell: $p^{\prime 2}_E=-M^{\prime 2}$. The
dimensionless variable $w_E$ is defined as
$w_E=v_E\cdot p^\prime_E/M^\prime=-w$.

The first integral appears in the calculation of
heavy-to-light form factors,
the second one in the calculation of the heavy-to-heavy case.

Scaling all momentum variables in (A.1) by $\Lambda_{B_q}$ and (A.2)
by $\Lambda_{B_Q}$ and
using the Feynman parametrization
$$\frac{1}{A}=\int\limits_0^\infty d\alpha \exp(-\alpha A)$$ we have
\begin{eqnarray}
& &I_1(-M^{\prime 2}, w_E)=2\Lambda_{B_q}[6(1+R)]^4
\exp[-36m^2(1+R)-9M^{\prime 2}\alpha^2]\\[4mm]
&\times&\int\limits_0^\infty ... \int\limits_0^\infty
d\beta_1 ... d\beta_4 t^2(\beta)
\int\frac{d^4 k_E}{\pi^2}\int\frac{d^4 k^\prime_E}{\pi^2}\nonumber\\[4mm]
&\times&\exp\biggl[-3(1+R)(1+\beta_3+\beta_4)
\biggl(k^\prime_E+k_E\frac{\beta_3-\beta_4}{1+\beta_3+\beta_4}\biggr)^2
\biggr]
\nonumber\\[4mm]
&\times&\exp\biggl[-3(1+R)t(\beta)(1+\beta_2)
\biggl(k_E+\frac{v_E\beta_1+p^\prime_Er_1}
{1+\beta_2}\biggr)^2\biggr]
\nonumber\\[4mm]
&\times&\exp\biggl[-\frac{3(1+R)t(\beta)}{1+\beta_2}
(\beta_1+M^\prime\beta_2-\bar\Lambda(1+\beta_2))^2
+6M^\prime(1+R)t(\beta)(w_E+1)\frac{\beta_1\beta_2}{1+\beta_2}\biggr]
\nonumber\\[4mm]
&\times&\exp\biggl[\frac{18M^\prime\alpha}{1+\beta_2}
(w_E\beta_1-M^\prime r_2)-12m^2(1+R)
\frac{(\beta_3-\beta_4)^2}{1+\beta_3+\beta_4}
-3(1+R)t(\beta)(4m^2-\bar\Lambda^2)]\nonumber\\[4mm]
&\times&
\exp\biggl[-3(1+R)t(\beta)\beta_2(m^2_s-(M^\prime-\bar\Lambda)^2)\biggr]
\nonumber
\end{eqnarray}
\begin{eqnarray}
& &I_2(w_E)=4\Lambda_{B_Q}^2 12^4\exp[-72m^2]
\int\limits_0^\infty ... \int\limits_0^\infty
d\beta_1 ... d\beta_4 t^2(\beta)
\int\frac{d^4 k_E}{\pi^2}\int\frac{d^4 k^\prime_E}{\pi^2}\\[4mm]
&\times&\exp\biggl[-6(1+\beta_3+\beta_4)
\biggl(k^\prime_E+k_E\frac{\beta_3-\beta_4}{1+\beta_3+\beta_4}\biggr)^2
\biggr]
\nonumber\\[4mm]
&\times&\exp\biggl[-6t(\beta)(1+\beta_2)
(k_E+v_E\beta_1+v^\prime_E\beta_2)^2\biggr]
\nonumber\\[4mm]
&\times&\exp\biggl[-6t(\beta)(\beta_1+\beta_2-\bar\Lambda)^2
+12t(\beta)(w_E+1)\beta_1\beta_2\biggr]
\nonumber\\[4mm]
&\times&\exp\biggl[-24m^2
\frac{(\beta_3-\beta_4)^2}{1+\beta_3+\beta_4}
-6t(\beta)(4m^2-\bar\Lambda^2)]\biggr]
\nonumber
\end{eqnarray}

\vspace*{0.5cm}
\noindent The notation is as follows:
\begin{eqnarray*}
& &R=\frac{\Lambda_{B_q}^2}{\Lambda_{B_Q}^2},\hspace*{.5cm}
t(\beta)=\frac{3+4(\beta_3+\beta_4)+4\beta_3\beta_4}{1+\beta_3+\beta_4},
\\[4mm]
& &r_1=\beta_2+\frac{3\alpha}{(1+R)t(\beta)},
\hspace*{.5cm}r_2=\beta_2+\frac{3\alpha}{2(1+R)t(\beta)}
\end{eqnarray*}

After a change of variables for $k^\prime_E$,  $k_E$
and integratious we arrive at
\begin{eqnarray}
& &I_1(-M^{\prime 2}, -w)=32\Lambda_{B_q}
\exp[-36m^2_q(1+R)-9m^{\prime 2}\alpha^2]\\[4mm]
&\times&\int\limits_0^\infty ... \int\limits_0^\infty
\frac{d\beta_1 ... d\beta_4}{(1+\beta_3+\beta_4)^2(1+\beta_2)^2}
\exp\biggl[-3(1+R)t(\beta)\beta_2(m^2_s-(M^\prime-\bar\Lambda)^2)\biggr]
\nonumber\\[4mm]
&\times&\exp\biggl[-\frac{3(1+R)t(\beta)}{1+\beta_2}
(\beta_1+M^\prime\beta_2-\bar\Lambda(1+\beta_2))^2
-6M^\prime(1+R)t(\beta)(w-1)\frac{\beta_1\beta_2}{1+\beta_2}\biggr]
\nonumber\\[4mm]
&\times&\exp\biggl[-\frac{18M^\prime\alpha}{1+\beta_2}
(w\beta_1+M^\prime r_2)-12m^2(1+R)
\frac{(\beta_3-\beta_4)^2}{1+\beta_3+\beta_4}
-3(1+R)t(\beta)(4m^2-\bar\Lambda^2)\biggr]
\nonumber
\end{eqnarray}
\begin{eqnarray}
& &I_2(-w)=64\Lambda_{B_Q}^2\exp[-72m^2]
\int\limits_0^\infty ... \int\limits_0^\infty
\frac{d\beta_1 ... d\beta_4}{(1+\beta_3+\beta_4)^2}\\[4mm]
&\times&\exp\biggl[-6t(\beta)(\beta_1+\beta_2-\bar\Lambda)^2
-12t(\beta)(w-1)\beta_1\beta_2\biggr]
\nonumber\\[4mm]
&\times&\exp\biggl[-24m^2
\frac{(\beta_3-\beta_4)^2}{1+\beta_3+\beta_4}
-6t(\beta)(4m^2-\bar\Lambda^2)\biggr]
\nonumber
\end{eqnarray}

\newpage
\begin{figure}[htbp]
\begin{center}
{\bf
\vspace*{-60mm}
\mbox{\epsfysize=11cm\epsffile{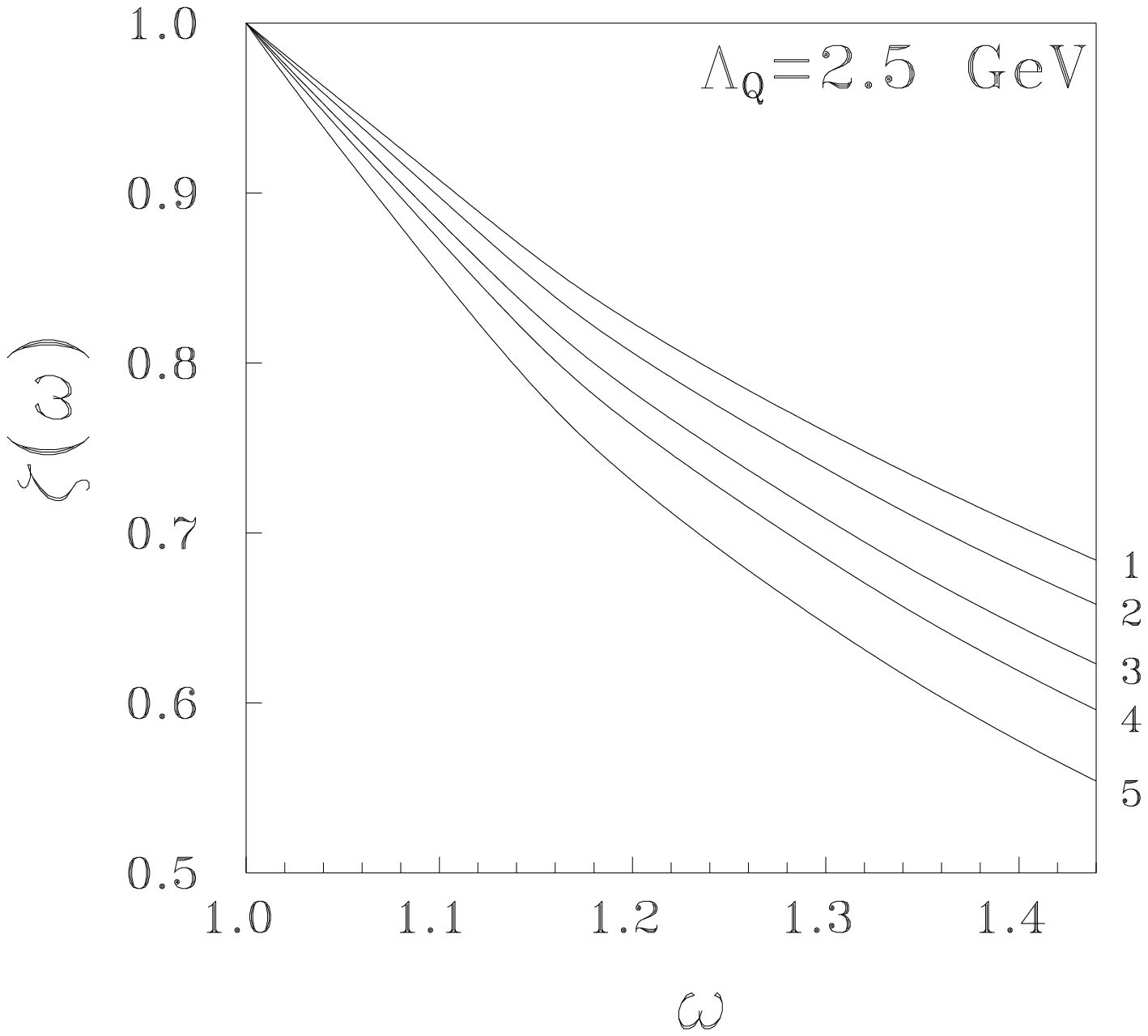}}
\mbox{\epsfysize=11cm\epsffile{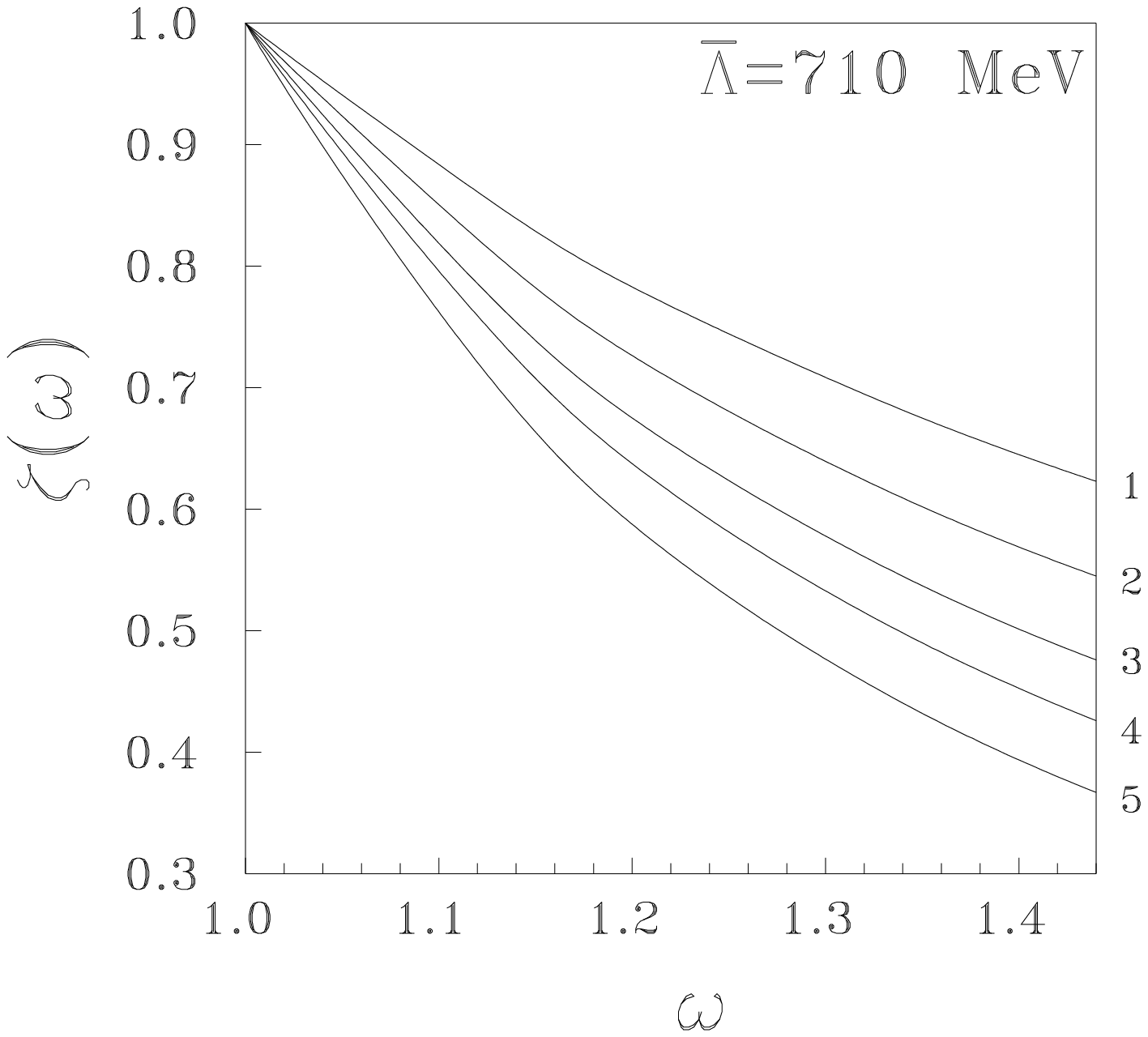}}
}
\end{center}
\end{figure}
\hspace*{1.2cm}{\bf Fig.3} $\zeta(\omega)$ function
\hspace*{4.2cm}{\bf Fig.4} $\zeta(\omega)$ function
\\
\hspace*{2cm}
1. $\bar\Lambda$=600 MeV \hspace*{4.5cm}   1. $\Lambda_Q$=2.5 GeV\\
\hspace*{2cm}
2. $\bar\Lambda$=650 MeV \hspace*{4.5cm}   2. $\Lambda_Q$=2.0 GeV\\
\hspace*{2cm}
3. $\bar\Lambda$=710 MeV \hspace*{4.5cm}   3. $\Lambda_Q$=1.7 GeV\\
\hspace*{2cm}
4. $\bar\Lambda$=750 MeV \hspace*{4.5cm}   4. $\Lambda_Q$=1.5 GeV\\
\hspace*{2cm}
5. $\bar\Lambda$=800 MeV \hspace*{4.5cm}   5. $\Lambda_Q\equiv\Lambda_q$=1.25 GeV\\

\begin{figure}[htbp]
\begin{center}
{\bf
\vspace*{-66mm}
\mbox{\epsfysize=11cm\epsffile{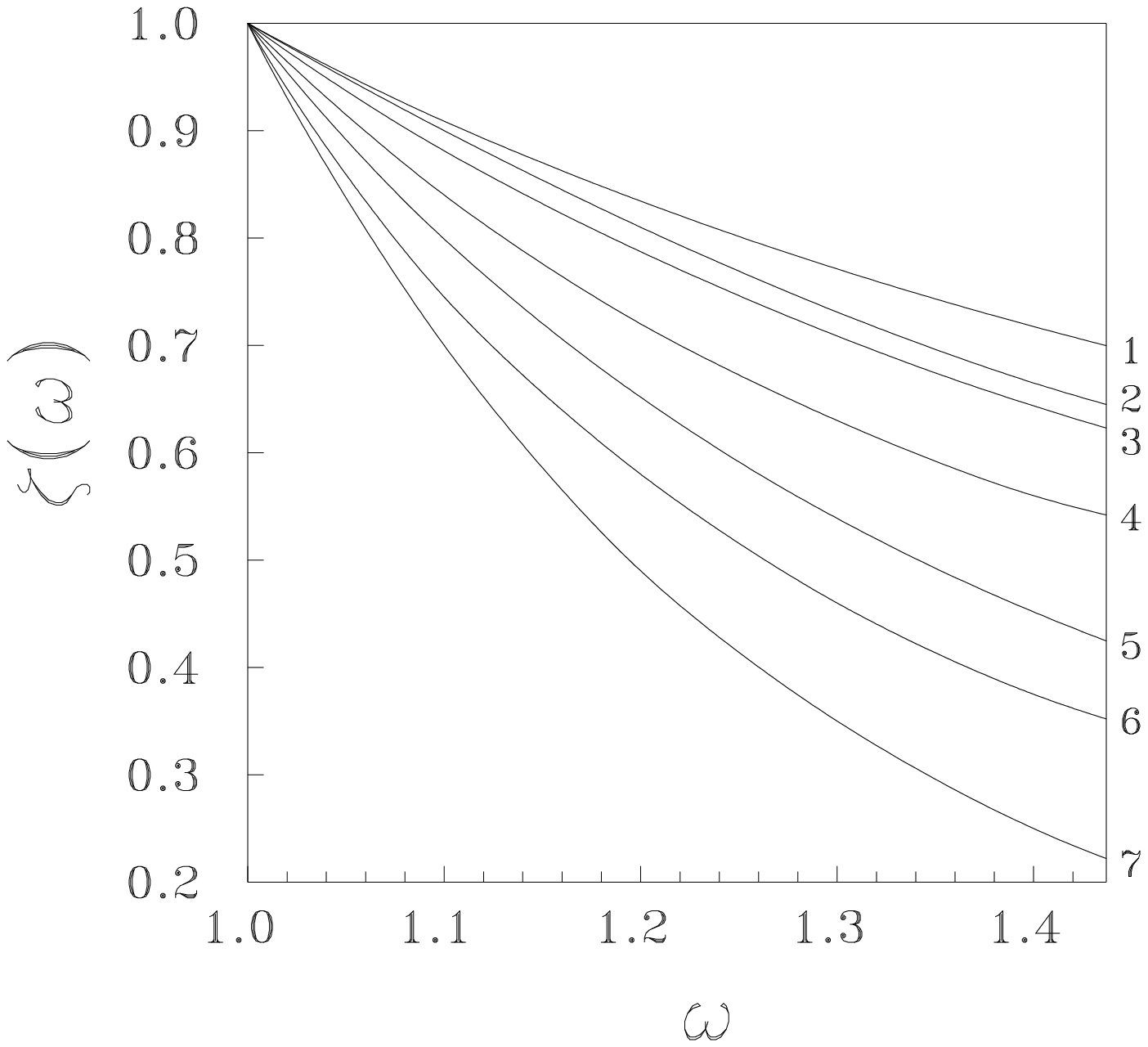}}
}
\end{center}
\end{figure}
\hspace*{6cm}{\bf Fig.5} $\zeta(\omega)$ function
\\
\hspace*{4cm}
1. SQM (Ref.\cite{Mark2}), 2. QCD SR (Ref.\cite{Grozin}), \\
\hspace*{4cm}
3. Our result ($\bar\Lambda$=710 MeV, $\Lambda_Q$=2.5 GeV), \\
\hspace*{4cm}
4. Dipole (Ref.\cite{DESY}), 5. MIT Bag (Ref.\cite{Zalewski}), \\
\hspace*{4cm}
6. IMF (Ref.\cite{DESY}), 7. IMF (Ref.\cite{Kroll})

\newpage
\begin{figure}[htbp]
\begin{center}
{\bf
\vspace*{-60mm}
\mbox{\epsfysize=11cm\epsffile{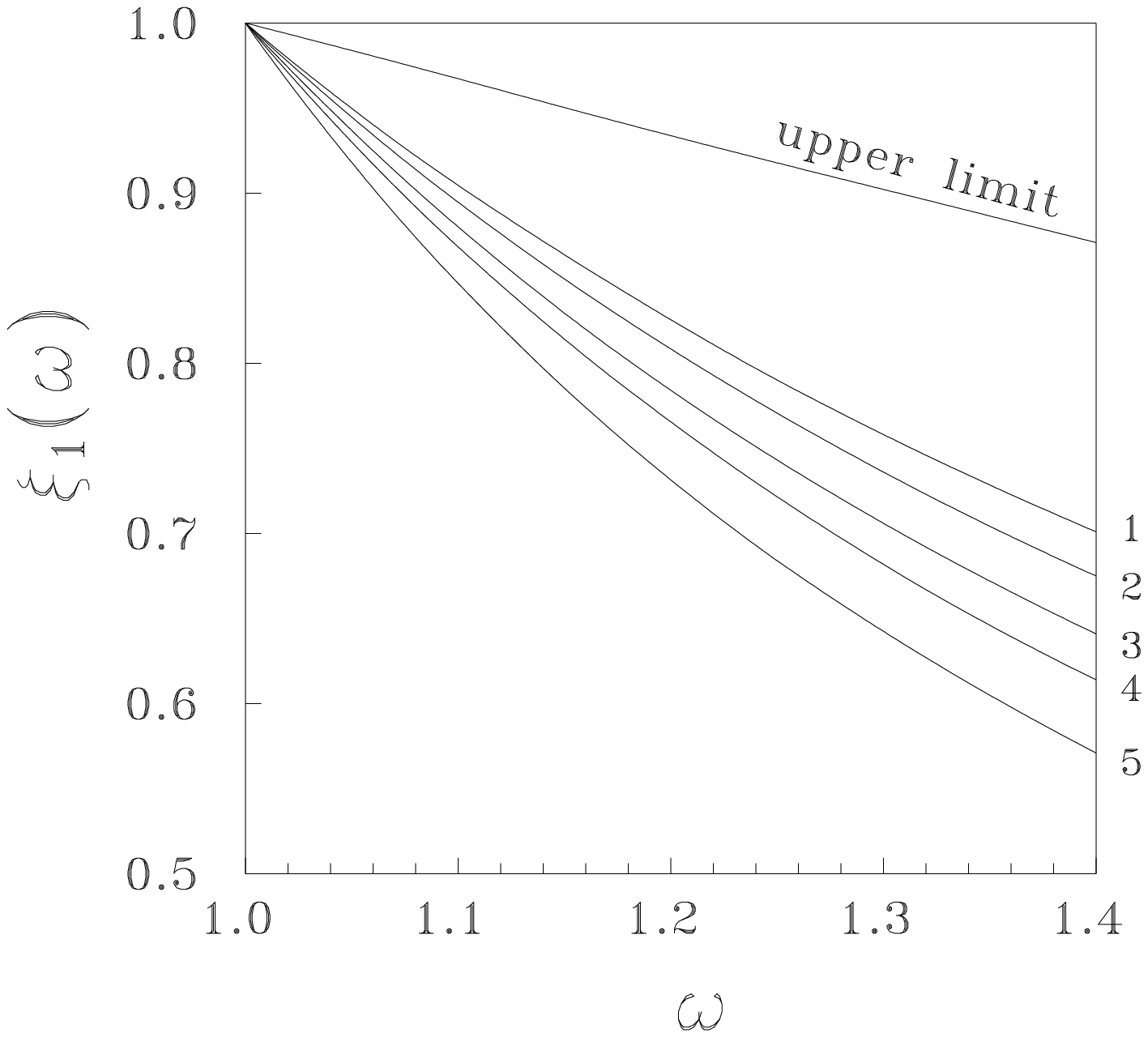}}
\mbox{\epsfysize=11cm\epsffile{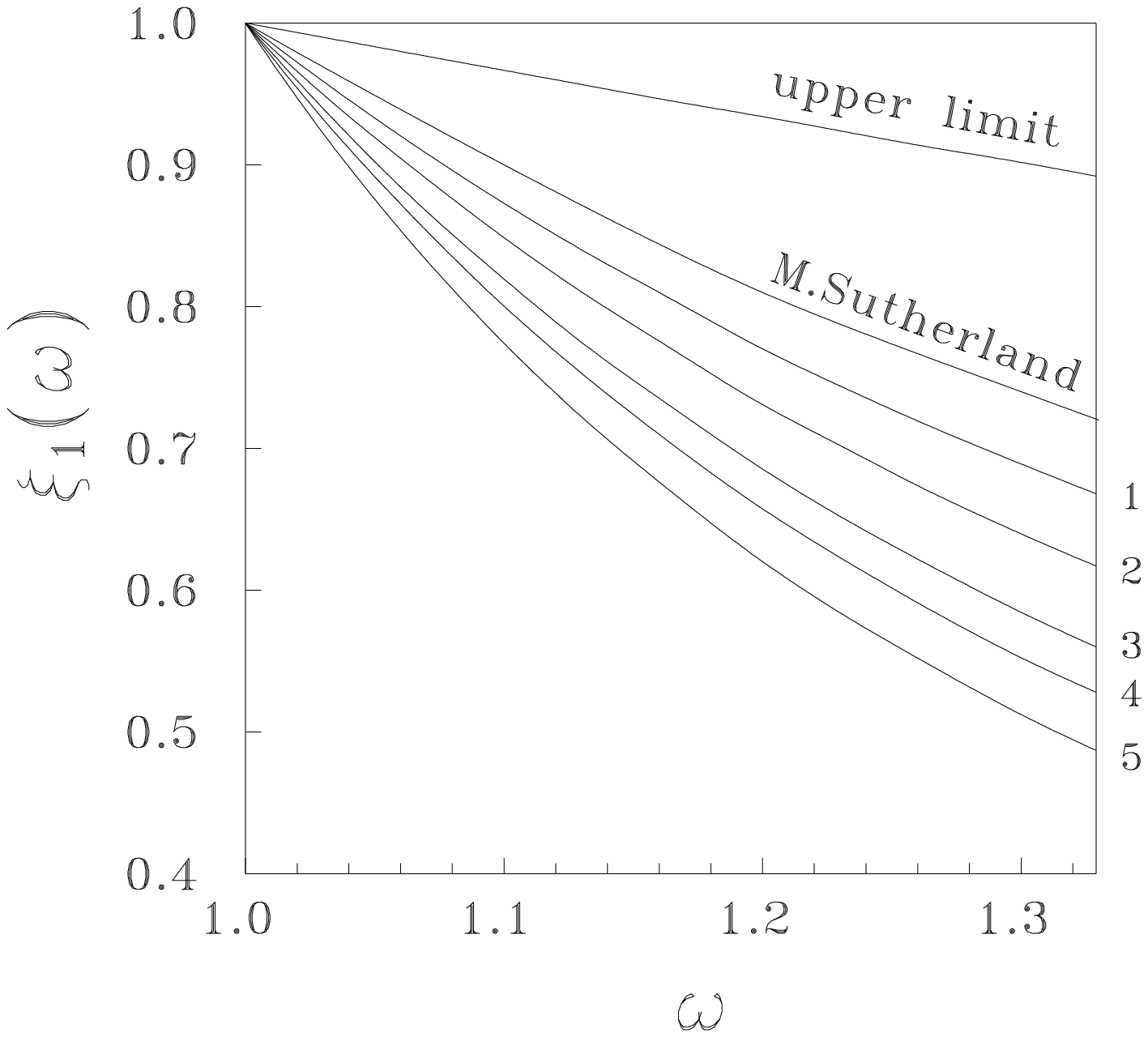}}
}
\end{center}
\end{figure}
\hspace*{0.7cm}{\bf Fig.6} $\xi_1(\omega)$ function ($\Sigma_b$-decay)
\hspace*{1.3cm} {\bf Fig.7} $\xi_1(\omega)$ function ($\Omega_b$-decay)
\\
\hspace*{2.5cm}
1. $\bar\Lambda$=600 MeV \hspace*{4.8cm}   1. $\bar\Lambda_{ss}$=800 MeV\\
\hspace*{2.5cm}
2. $\bar\Lambda$=650 MeV \hspace*{4.8cm}   2. $\bar\Lambda_{\{ss\}}$=900 MeV\\
\hspace*{2.5cm}
3. $\bar\Lambda$=710 MeV \hspace*{4.8cm}   3. $\bar\Lambda_{\{ss\}}$=1000 MeV\\
\hspace*{2.5cm}
4. $\bar\Lambda$=750 MeV \hspace*{4.8cm}   4. $\bar\Lambda_{\{ss\}}$=1050 MeV\\
\hspace*{2.5cm}
5. $\bar\Lambda$=800 MeV \hspace*{4.8cm}   5. $\bar\Lambda_{\{ss\}}$=1100 MeV\\

\newpage

\begin{figure}[htbp]
\begin{center}
{\bf
\vspace*{-60mm}
\mbox{\epsfysize=11cm\epsffile{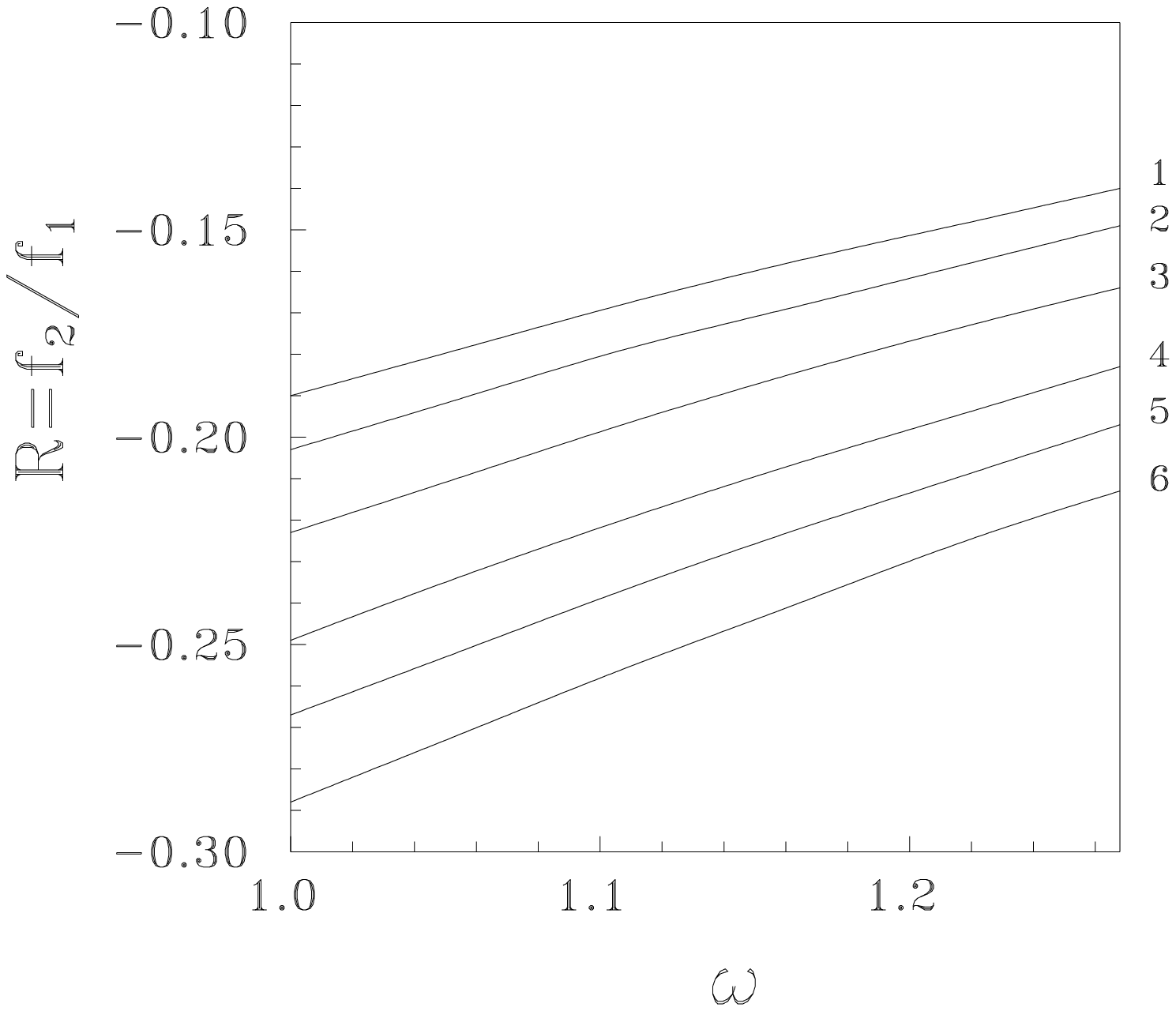}}
\mbox{\epsfysize=11cm\epsffile{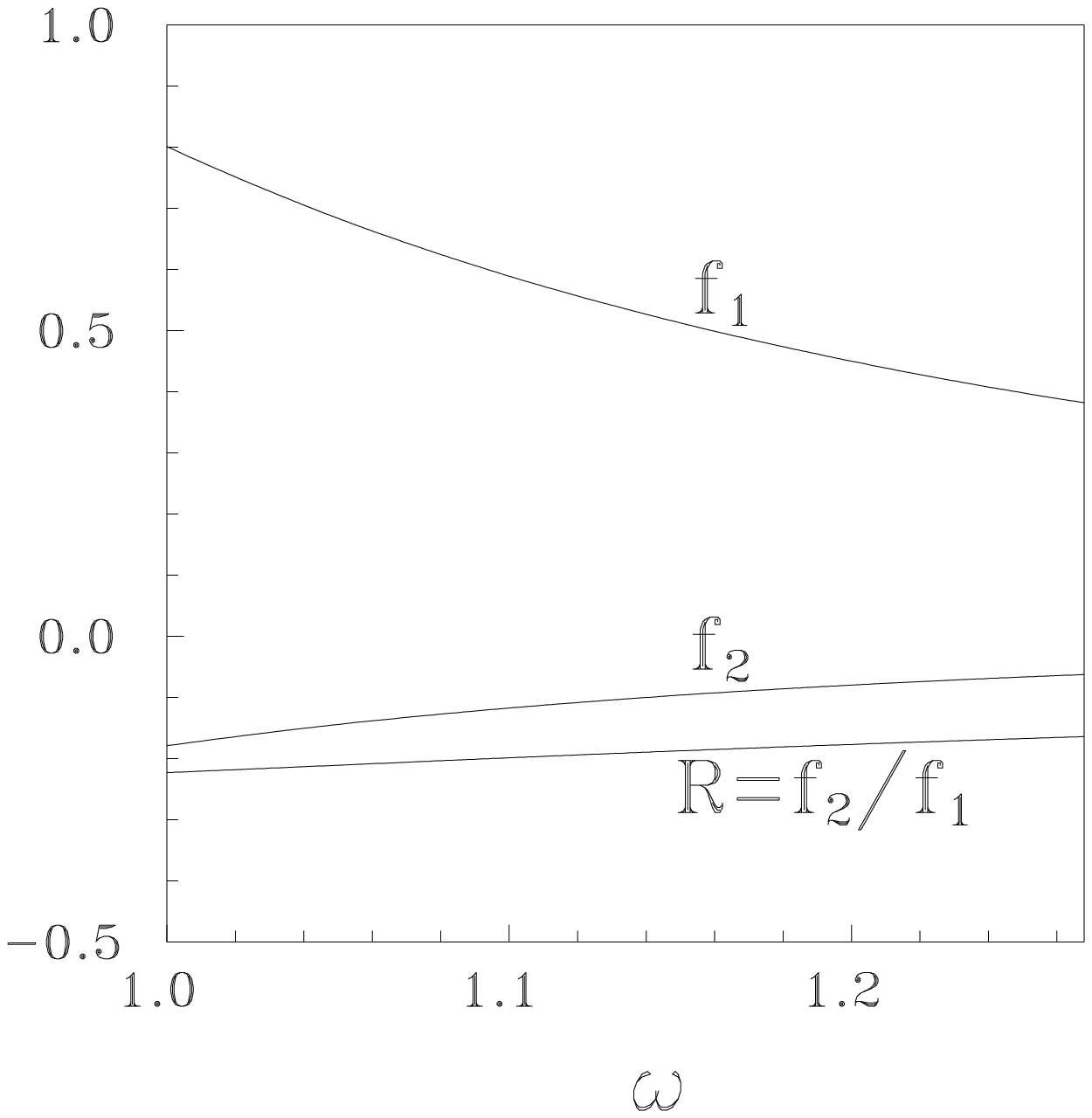}}
}
\end{center}
\end{figure}
{\bf Fig.10} Form factor ratious $R=f_2/f_1$
\hspace*{1cm} {\bf Fig.11} Form factors and their ratio\\
\hspace*{2cm}for $\Lambda^+_c\to\Lambda^0+e^+\nu$-decay\hspace*{3.5cm}
for $\Lambda^+_c\to\Lambda^0+e^+\nu$-decay
\\
\hspace*{2.5cm}
1. $\bar\Lambda$=650 MeV\\
\hspace*{2.5cm}
2. $\bar\Lambda$=710 MeV\\
\hspace*{2.5cm}
3. $\bar\Lambda$=725 MeV\\
\hspace*{2.5cm}
4. $\bar\Lambda$=750 MeV\\
\hspace*{2.5cm}
5. $\bar\Lambda$=775 MeV\\
\hspace*{2.5cm}
6. $\bar\Lambda$=800 MeV\\

\newpage
\begin{figure}[htbp]
\begin{center}
\vspace*{-70mm}
{\bf
\mbox{\epsfysize=14cm\epsffile{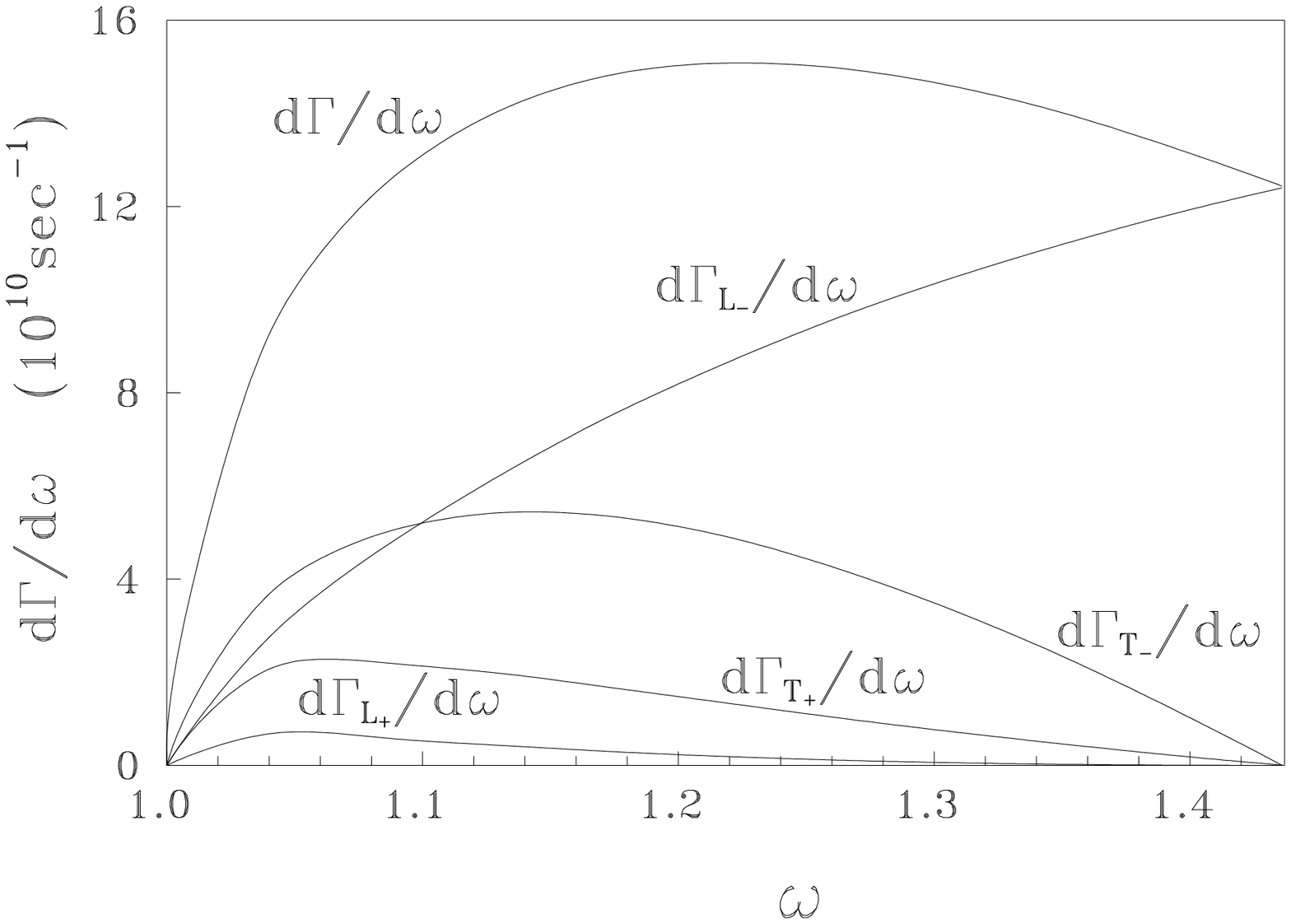}}}
\\
\vspace*{2cm}
\mbox{\hspace*{-0.3cm}{\bf Fig.8} Differential Distribution}
\\
\vspace*{-60mm}
{\bf
\mbox{\epsfysize=14cm\epsffile{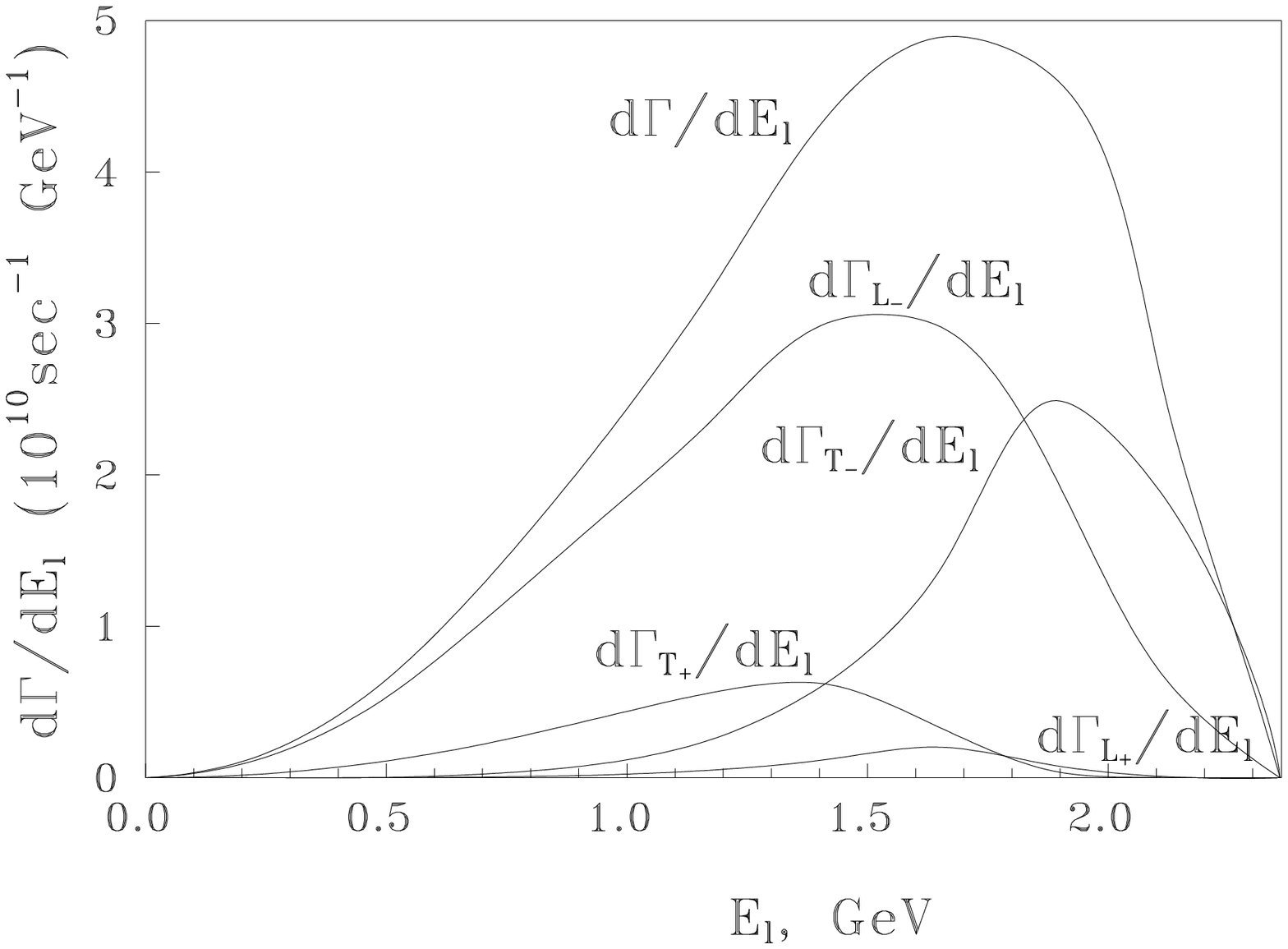}}}
\\
\vspace*{2cm}
\mbox{\hspace*{-0.3cm}{\bf Fig.9} Leptonic Spectrum}
\end{center}
\end{figure}
\end{document}